\documentclass[floatfix,aps,twocolumn,floats,nofootinbib,pre]{revtex4-2}
\usepackage{graphics,graphicx,epsfig}
\usepackage{epsf,epstopdf,wrapfig}
\usepackage{amsmath,amssymb}
\usepackage[dvipsnames]{xcolor}
\usepackage{tikz}
\usepackage{dsfont}

\usepackage{newfloat}
\usepackage{algorithm}
\usepackage{algpseudocode}
\usepackage{mathtools}
\usepackage{hyperref}
\usepackage[normalem]{ulem}

\begin{document}

\title{A Markovian dynamics for \emph{C. elegans} behavior across scales}

\author{Antonio C. Costa\textsuperscript{a}}\thanks{Current address: Laboratoire de Physique de l’Ecole normale supérieure, ENS, Université PSL, CNRS, Sorbonne Université, Université Paris Cité, F-75005 Paris, France}
\author{Tosif Ahamed\textsuperscript{b,c}}
\author{David Jordan\textsuperscript{d}}
\author{Greg J. Stephens\textsuperscript{a,c}}

\affiliation{$^a$Department of Physics and Astronomy, Vrije Universiteit Amsterdam, 1081HV  Amsterdam, The Netherlands\\
$^b$HHMI Janelia Research Campus, Ashburn, VA, USA\\
$^c$Biological Physics Theory Unit, OIST Graduate University, Okinawa 904-0495, Japan \\
$^d$Department of Biochemistry, University of Cambridge, Cambridge, UK 
}

\begin{abstract} 
How do we capture the breadth of behavior in animal movement, from rapid body twitches to aging? Using high-resolution videos of the nematode worm \emph{C. elegans}, we show that a single dynamics connects posture-scale fluctuations with trajectory diffusion, and longer-lived behavioral states. We take short posture sequences as an instantaneous behavioral measure, fixing the sequence length for maximal prediction. Within the space of posture sequences we construct a fine-scale, maximum entropy partition so that transitions among microstates define a high-fidelity Markov model, which we also use as a means of principled coarse-graining. We translate these dynamics into movement using resistive force theory, capturing the statistical properties of foraging trajectories.  Predictive across scales, we leverage the longest-lived eigenvectors of the inferred Markov chain to perform a top-down subdivision of the worm's foraging behavior, revealing both ``runs-and-pirouettes'' as well as previously uncharacterized finer-scale behaviors. We use our model to investigate the relevance of these fine-scale behaviors for foraging success, recovering a trade-off between local and global search strategies. 
\end{abstract}


\maketitle

\section*{Significance Statement} 
Complex phenotypes, such as an animal's behavior, generally depend on an overwhelming number of processes that span a vast range of scales. While there is no reason that behavioral dynamics permit simple models, by subsuming inherent nonlinearities and memory into maximally-predictive microstates, we find one for \emph{C. elegans} foraging. The resulting ``Markov worm'' is effectively indistinguishable from real worm motion across a range of timescales, and we can decompose our model dynamics both to recover and discover behavioral states. Finally, we connect postures to trajectories, illuminating how worms explore the environment in different behavioral states.

\section*{Introduction}

From molecular motors contracting muscles, to neurons processing an ever changing environment, or the large-scale diffusion of hormones and other neuromodulatory chemicals, animal behavior arises from biological activity across innumerable spatial and temporal scales.  With an instantaneous snapshot of all of these variables, the future behavioral state of the animal would be uniquely defined, a biological setting for the demon of Laplace (see e.g.~\cite{Weinert2016}). Of course, such an approach is practically unrealizable. We are limited to a much smaller set of observations and the unobserved degrees of freedom will generally induce non-Markovianity, or memory, to the dynamics of the variables that we do measure \cite{Mori1965,Zwanzig1973}. In animal behavior, interpretations of this memory guide our understanding of the complexity of the process \cite{Berman2018,Bialek2023_PRL}. But what if we could use our observations to construct memory-full state variables that admit predictive, yet minimal-memory dynamics \cite{Costa2023,Rupe2022}? 

The construction of such dynamics appears daunting. We may even conclude that this is impossible if it were not for the fact that it is done routinely in physical systems. Indeed, it is often the case that a subset of observable functions is enough to capture behavior at a particular scale. Hydrodynamics, for example, can be formulated with effective variables such as fluid velocity, density, or temperature: their memory only coming from the previous state. In behavior, we expect the emergent reconstructed dynamics to be generally high-dimensional in order to account for the multitude of unobserved mechanisms. Yet our approach also suggests a principled coarse-graining. Since the dynamics of the reconstructed states are Markovian, the emergent timescales of the (nonlinear) system are naturally ordered by the eigenvalue spectrum of a \emph{linear} evolution operator, or transition matrix in the case of discrete states. The eigenvectors associated with the gaps in the spectrum indicate slow collective modes and provide natural targets for coarse-graining.  In the hydrodynamic example of $\sim 10^{23}$ interacting molecules, these modes are the effective variables.

Here we seek such Markov dynamics from the time series of posture in the foraging behavior of the nematode worm \emph{C. elegans}, an important model organism in genetics and neuroscience \cite{Brenner1974,Bargmann1993,Bargmann2013}. For both the worm and animals generally, the collection of high-resolution behavioral data has been greatly accelerated by advancing techniques for pose estimation via machine vision \cite{Pereira:2019te, Mathis:2018us,Hebert2021,Pereira2020,Mathis2020}, combined with computational and imaging improvements. Such measurement advances demand new behavioral understanding: analyses, models, and theory of posture-scale dynamics \cite{Aguilar2016,Berman2018,Brown2018}.

We implement a principled, generally-applicable framework which combines delay embedding with Markov modeling \cite{Costa2023}. In this approach, we seek to overcome the partial observability of behavioral dynamics; variables which influence behavior but are instantaneously hidden become apparent over time and Markov predictability provides the quantitative measure of a self-determined system.  Posture itself is a very complicated function of its underlying biological variables. In such situations, an initial expansion of dimensionality can simplify computations like function estimation and classification.  We thus trade the complex modeling of a low-dimensional time series for the simpler modeling of a much higher-dimensional state space: the encoding of the unobserved degrees of freedom through time delays drastically simplifies our theory, leading to a powerful yet simple description of the emergent nonlinear Markovian dynamics.

While Markov approaches have an extensive history, perhaps most familiarly in Markov Chain Monte Carlo sampling of \emph{equilibrium} distributions \cite{Landau2014}, substantially less attention has focused on a Markov encoding of actual \emph{dynamics}, especially with a large number of states. Importantly, we note that there is no guarantee that our approach will work; for example, the number of necessary delays may be computationally prohibitive. But even this ``failure'' would provide important information about the memory of the system.  On the other hand, if we are successful, we will be left with a finite set of observables that are approximately self-determined, measurable, and whose dynamics span the timescales that are relevant to the phenomena of interest.  Such observables are likely to be biologically meaningful.

We find state variables for worm behavior that exhibit Markovian evolution across the multiple timescales of \emph{C. elegans} foraging behavior: from fine-scale posture movements to ``run-and-pirouette'' strategies. Additionally, the macroscopic variables we reveal are not some baroque non-physical mathematical functions but rather correspond to interpretable behavioral motifs.  We rediscover canonical behaviors from the rich history of \emph{C. elegans} ethomics, as well as describe new ones.  Each of these motifs is associated with its own characteristic timescale, and with them we provide a new hierarchical subdivision of behavior.  We show how the dynamics of these macroscopic variables can be propagated through a model of the organisms physical interaction with the environment to accurately predict locomotion from posture. Finally, we dissect the function of these behavioral motifs by investigating their relation to the exploration and exploitation of food sources.

\section*{Short-time behaviors as maximally-predictive posture sequences} 

On a 2D agar plate, worms move by making dorsoventral bends along their bodies \cite{Croll1975}. At the shortest timescales $(\sim 1\, s)$ these traveling waves along the body give rise to forward, backward, and turning locomotion \cite{Stephens2008,Costa2019,Ahamed2021}.  We show here that this organization emerges naturally from short posture sequences, formally a delay embedding first introduced in the theory of dynamical systems \cite{Takens1981,Sugihara1990,Sauer1991,Stark1999,Stark2003}.

\begin{figure*}[ht!]
\centering
\includegraphics[width=.8\linewidth]{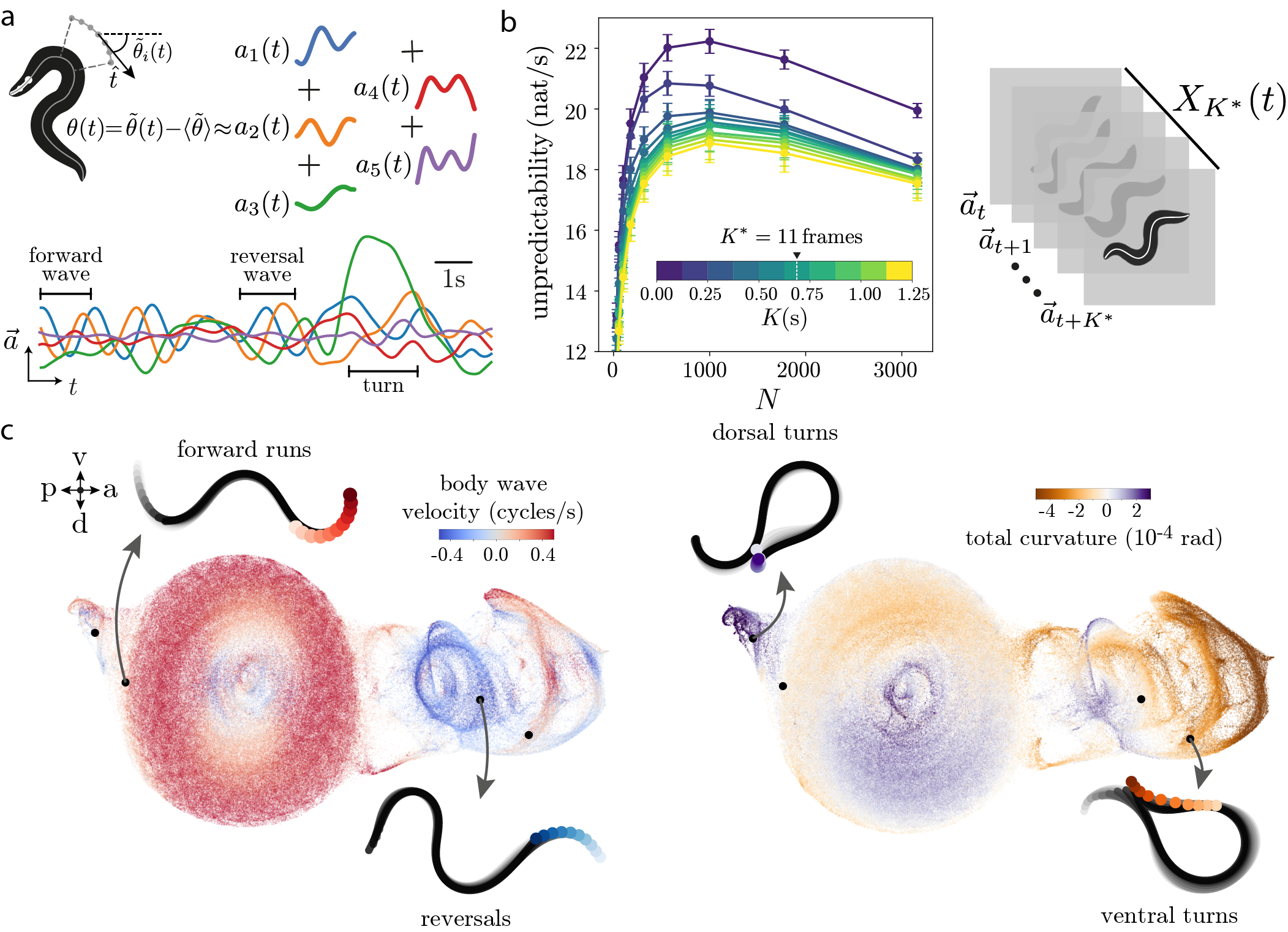}
\caption{{\bf Maximally-predictive posture sequences reveal the space of short-time behaviors.}
(a-left) We represent the posture at each frame $t$ using an ``eigenworm'' basis \cite{Stephens2008}. We extract the worm's centerline and measure the tangent angles between body segments $\tilde{\theta}_i(t)$. We then subtract the average angle to obtain a worm-centric representation $\theta = \tilde{\theta} - \langle \tilde{\theta}\rangle$, and project the mean-subtracted angles onto a set of eigenworms \cite{Stephens2008}, obtaining a 5D (eigenworm coefficient) time series $\vec{a}(t)$. 
(a-bottom) A segment of the time series: in food-free conditions, the short time scale behavior roughly consists of forward and reversal waves (clearly visible as oscillations in $a_1$ and $a_2$), as well as sharp turns (characterized by large amplitude $a_3$ \cite{Broekmans2016}). 
(b) Entropy rate as a function of the sequence length $K$ and number of partitions $N$. We partition each sequence space (indexed by length $K$) into N microstates using k-means clustering and compute the entropy rate of the resulting Markov chain. The curves collapse after $K\sim 0.5\,\text{s}$, indicating that the entropy rate is approximately constant meaning that there is no further gain in predictability by including more time delays. We choose $K^* = 11\,\text{frames} = 0.6875\,\text{s}$ to define a maximally predictive sequence space $X_{K^*}$. Error bars are bootstrapped standard deviations across worms.
(c) We visualize $X_{K^*}$ by projecting onto two-dimensions using UMAP \cite{McInnes2018}, and coloring each point by the body wave phase velocity $\omega = -\frac{1}{2\pi}\frac{d}{dt}\left[\text{tan}^{-1}(a_2/a_1)\right]$ \cite{Stephens2008} (left) and the overall body curvature (right) obtained by summing the tangent angles $\theta_i$ along the body $\gamma = \sum_i \theta_i$. Each point on this space corresponds to a $K^*$ sequence of postures (from light to dark colors), and different short-time behaviors naturally correspond to different regions on the projection.
}
\label{fig_1}
\end{figure*}

We employ a previously analyzed dataset \cite{Broekmans2016} composed of 35 minute recordings of 12 young-adult lab-strain N2 worms freely moving on an agar plate, sampled at $\delta t = 1/16 \,\text{s}$ (a total of 33,600 frames per worm). From high-resolution videos, we measure the worm's body posture using a rich low dimensional representation of the centerline, expressed as five ``eigenworm'' coefficients $\vec{a}=[a_1,a_2,a_3,a_4,a_5]\in\mathbb{R}^5$ \cite{Stephens2008,Broekmans2016}, Fig.\,\ref{fig_1}(a). Posture itself already carries important behavioral information, for example the joint distribution of the first two eigenworm modes is approximately circular, describing the phase of the crawling locomotory wave \cite{Stephens2008}.  But posture {\em dynamics} carries even more information, such as the direction of the wave which distinguishes forward crawls from reversals.

We capture posture dynamics through a maximally predictive sequence space \cite{Costa2023} constructed by stacking $K$ delays of the posture time series, and increasing $K$ until we have maximized the predictability of the resulting dynamics, as measured by the entropy rate, Fig.\,\ref{fig_1}(b-left). While other quantitative approaches to behavior also start with short dynamical motifs \cite{Berman2014a, Wiltschko2015} our work is distinguished by a focus on maximal predictability to determine the motif length. To estimate the entropy rate, we partition each sequence space (indexed by $K$) into $N$ microstates $s_i$ using k-means clustering so that the posture dynamics now appear as transitions between microstates, a Markov chain. We build the partition using all $12\times 33600$ observations so that the microstate labelling is consistent across worms. We approximate the entropy rate as that of the Markov chain inferred from each worm's symbolic sequence and choose the largest $N$ before finite size sampling reduces the estimated entropy\footnote{The entropy rate of the Markov chain should not be confused with the Kolmogorov-Sinai (KS) entropy, which is an intrinsic property of the dynamics. Indeed, the KS entropy can also be estimated using our approach, yielding accurate estimates that agree with the sum of positive Lyapunov exponents \cite{Costa2023,Ahamed2021}.}.  This ``maximum entropy'' partitioning (with $N^*=1000$ partitions) requires a large number of microstates, but enables our model to be maximally expressive within the limits of the data (which set the onset of finite size effects in the entropy estimation). After $K\approx 8\,\text{frames} = 0.5\,\text{s}$ and the entropy rate curves start to collapse, and we set $K^*=11\,\text{frames} = 0.6875 \,\text{s}$ to define the maximally-predictive sequence space $X_{K^*}$ of dimensions $5\times 11$, Fig.\,\ref{fig_1}(b-right); lengthening the sequences does not increase predictability, Fig.\,S1(a).  While Fig.\,S1(a) suggests that $K^*=6\,\text{frames}$ is enough to maximize predictability, we choose a slightly higher value of $K^*$ to ensure that we fully resolve the predictive information lying in past posture measurements.

We visualize the high-dimensional $X_{K^*}$ by projecting into two dimensions using the UMAP manifold learning algorithm \cite{McInnes2018} (see SI Appendix: Methods), Fig.\,\ref{fig_1}(c). Color-coding according to the worm's body wave phase velocity $\omega = -\frac{1}{2\pi}\frac{d}{dt}\left[\text{tan}^{-1}(a_2/a_1)\right]$, \cite{Stephens2008} Fig.\,\ref{fig_1}(c, left), and overall curvature (obtained by summing the tangent angles along the body), $\gamma = \sum_i \theta_i$, reveals that distinct short-time behavioral motifs, corresponding to forward, reversal, and turning movements, naturally correspond to different regions of the maximally predictive state space, Fig.\,\ref{fig_1}(c, right). In other words, while the instantaneous posture itself $\vec{a}(t)$ is not enough to disentangle different behaviors, a point in the sequence space $X_{K^*}(t)$ uniquely corresponds to a particular short-time behavioral motif.

\section*{The Markov Worm}

Our combination of sequence embedding and partitioning naturally results in a symbolic dynamics where each state probabilistically transitions to a new state \cite{Costa2023}. These dynamics are described through the master equation
\begin{align}
p_j(t+\tau)=\sum \limits_{i} P_{ij}(\tau)\,p_i(t),  
\label{eq:MarkovChain}
\end{align}
where $p_i(t)$ is the probability of observing state $s_i$ at time $t$, and $P_{ij}(\tau)$ is a transition matrix constructed by counting transitions between microstates $s_i(t)$ and $s_j(t+\tau)$ after a delay $\tau$.  We note that $P$ is a stochastic matrix so that each transition probability $P_{ij} \geq 0$ and $\sum_j P_{ij}=1$ for all microstates $i$. The waiting time $\tau$ is a free parameter which can be as short as $\delta t$, as we used previously for the maximally-predictive embedding, and we will leverage this freedom to emphasize long-lived aspects of the worm's movement. We have thus transformed the behavioral dynamics into a (generally non-equilibrium) Markov chain with many states, and where each state itself carries short-time dynamical information.

\begin{figure*}[ht!]
\begin{center}
\includegraphics[scale=1]{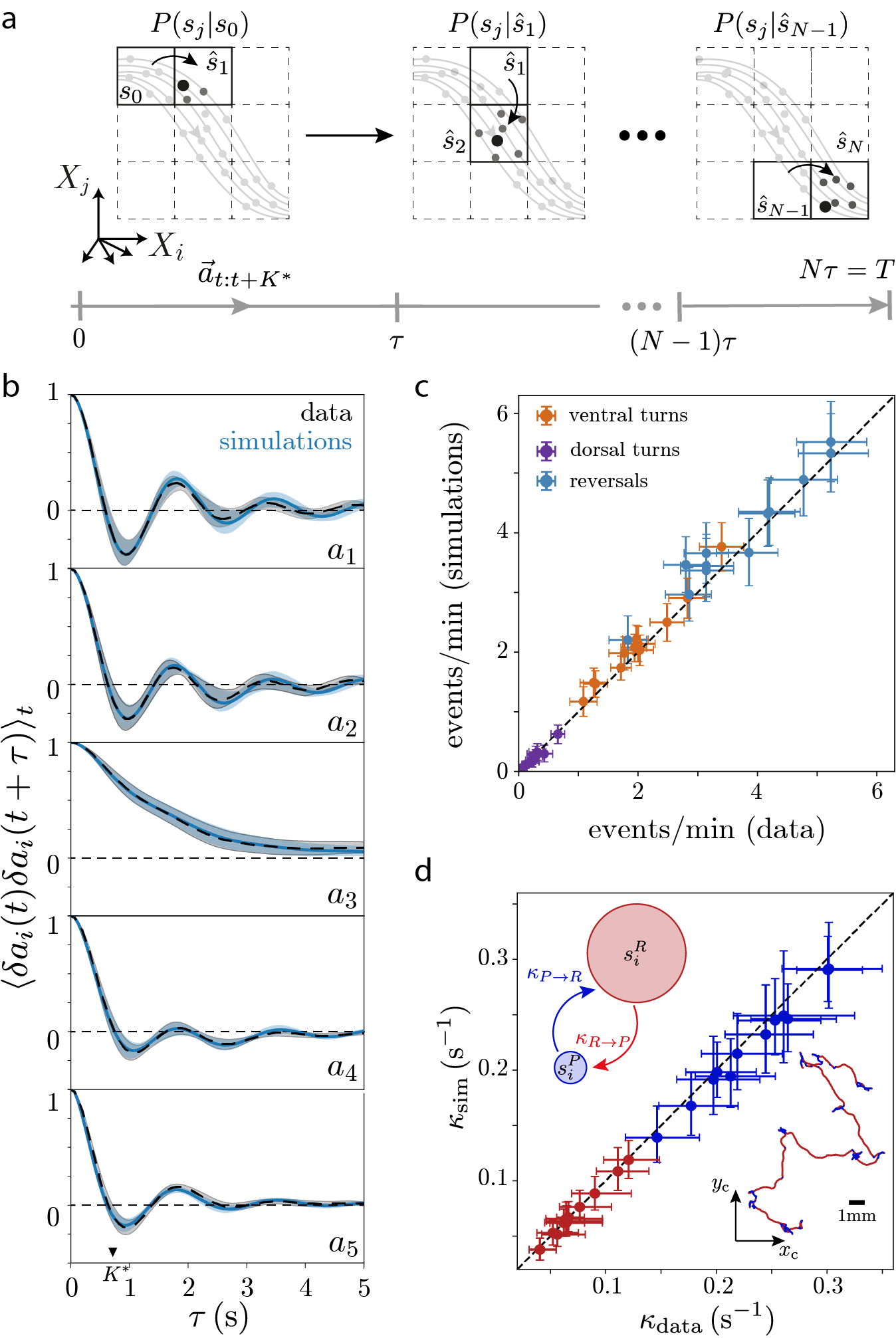}
\caption{{\bf A Markov model captures posture dynamics across timescales.}
(a) Schematic of the simulation method. Starting from the initial microstate of each worm $s_0$, the next microstate is obtained by sampling from $P_{0j}(\tau^*)$. We add new microstates in the same fashion, resulting in a symbolic sequence of length $N=T\delta t/\tau^*$. Within each microstate, we randomly choose a point in the maximally-predictive sequence space, $X_{K^*}$, and use this point to identify the associated sequence of body postures $\vec{a}_{t:t+K^*}$.
(b) The simulated posture dynamics accurately predicts the autocorrelation function of the ``eigenworm'' coefficient time series. Shaded regions correspond to 95\% confidence intervals of the estimate of the mean autocorrelation function bootstrapped across worms.
(c) The simulated posture dynamics accurately predicts the average rate of behavioral events across worms. We estimate the number of reversals, ventral turns and dorsal turns per unit time, and compare the result obtained from the data with simulations of each of the 12 recorded worms, finding excellent agreement. Error bars correspond to bootstrapped 95\% confidence intervals on the estimates of the mean.
(d) On longer time scales, worms transition between relatively straight ``runs'' and abrupt reorientations through ``pirouettes'' \cite{Pierce-Shimomura1999}, which are combinations of reversals and turns (inset, bottom right).  We assign each microstate to either a ``run'' or a ``pirouette'' by leveraging the inferred transition matrix to identify long-lived stereotyped sequences of posture movements (see section {\bf COARSE-GRAINING BEHAVIOR THROUGH ENSEMBLE DYNAMICS}). We then estimate the mean transition rates from the run to the pirouette state $\kappa_{R\rightarrow P}$ (red) and from the pirouette to the run state $\kappa_{P\rightarrow R}$ (blue) for each of the 12 recorded worms. Data error bars are 95\% confidence intervals of the mean bootstrapped across run and pirouette segments, simulation error bars are 95\% confidence intervals of the mean bootstrapped across 1000 simulations. 
}
\label{fig_2}
\end{center}
\end{figure*}

As described in the previous section, for the worm's posture dynamics, we set the number of microstates as $N^*=1000$ so as to maximize the amount of information with respect to the partitioning, Fig.\,\ref{fig_1}(b). We then set the transition time as $\tau^* = 0.75\,\text{s}$ so that the relaxation times of the long-lived dynamics are approximately independent of $\tau$, as rigorously true in a Markov process (see Fig.\,S1(b,c) and SI Appendix: Methods).  In practice, from the $\text{33,600}-K^*-\tau^*$ observations of each worm, we count how many times state $s_i$ reaches state $s_j$ in a timescale $\tau^*$. Since $\tau^*$ is much shorter than the mixing time of the dynamics, most transitions occur locally in $X_{K^*}$, so we expect that the number of nonzero transition probabilities is much smaller than $N^* \times N^*$. Indeed we find that from each $s_i$ the dynamics only reaches $\sim 10$ other states within $\tau^*$  yielding transition matrices with $\approx 10,000$ entries, Fig.\,S4. Despite the conceptual simplicity of our Markov chain model, Eq.\ref{eq:MarkovChain}, we next show that it accurately predicts \emph{C. elegans} foraging behavior across scales, from fine-scale posture movements to long time scale transitions between behavioral states, Fig.\,\ref{fig_2},S4.

\subsection*{Predicting behavior across scales}

Starting from the initial state of an individual worm, which corresponds to a discrete microstate $s_i(t=0)$ of the $K^*$ space of posture sequences, we simulate symbolic sequences by sampling from the conditional probability distribution $P^w(s_j|\hat{s}_i)$, Fig.\,\ref{fig_2}(a), where $\hat{s}_i$ is the current microstate, $s_j$ are all possible future microstates after a time scale $\tau^*$ and $P^w(s_j|\hat{s}_i)$ is the $i$-th row of the transition matrix inferred for worm $w$. The result is a sequence of microstates with the same duration as the worm trajectories, but with a sampling time $\delta t = \tau^*$. From each microstate we can then obtain a nearly continuous time series of ``eigenworm'' coefficients $\{\vec{a}(t)\}$ through the sequence of $K^*$ postures in each state $X_{K^*}$ (note that $K^*$ and $\tau^*$ are quite close in this case). These dynamics are effectively diffusive in the space of posture sequences: hopping between microstates according to the Markov dynamics, followed by random selection from the set of posture sequences ${X^i_{K^*}}$ within each visited microstate $s_i$. The posture time series generated through this procedure are nearly indistinguishable from the data, Fig.\,S2 and SI Movie 1. Quantitatively, the autocorrelation functions of the simulated time series, Fig.~\ref{fig_2}(b), S4(a), capture the correlations observed in the data, and the distribution of mode coefficients agrees with the steady-state distribution, Fig.\,S3. 

In addition to the finescale posture dynamics, our model also predicts the rate at which forward movements are interrupted by biologically relevant behaviors \cite{Zhao2003} such as reversals, dorsal turns or ventral turns (identified by thresholding the body wave phase velocity \cite{Stephens2008} and the overall body curvature, see SI Appendix: Methods), Fig.\,\ref{fig_2}(c), S4(b). 

At larger spatio-temporal scales the foraging random-walk can be coarsely split into forward ``runs'' interrupted by ``pirouettes'', a single or sequence of sharp centroid turns, which are used by the worm to reorient itself between longer-lived runs \cite{Pierce-Shimomura1999}.  We note that the term ``pirouette'' was originally introduced to broadly describe the reversals, deep turns and combinations of reversals and deep turns seen during chemotaxis \cite{Pierce-Shimomura1999}. This definition has also been applied to the worm's navigation behavior in food-free environments \cite{Gray2005}, similar conditions to the recordings described here. Here we identify long-lived behavioral states from posture dynamics by using the inferred transition matrix to find sets of movements that typically occur in a sequence (we call these  ``macrostates'', see the following section COARSE-GRAINING BEHAVIOR THROUGH ENSEMBLE DYNAMICS). At the two-state level, the macrostates split the centroid trajectory into relatively-straight paths and more abrupt reorientations, and we thus identify these macrostates as the posture-scale expression of  ``runs'' and ``pirouettes'' described above, Fig.\,\ref{fig_2}(d-inset). We estimate the kinetic transition rates from runs-to-pirouettes $\kappa_{R\rightarrow P}$ and from pirouettes-to-runs $\kappa_{P\rightarrow R}$ and find close agreement between data and simulations across worms, Fig.\,\ref{fig_2}(d),S4(c).  

\subsection*{Posture to Path}

Our Markov model on posture sequences is remarkably powerful at predicting worm behavior from an egocentric point of view. However, to understand how such dynamics relate to biological function, we need to connect posture dynamics to motility in the 2D space. With such a bridge we could, for the first time, connect the neuromechanical control of posture with movement strategies such as optimal search. Adding posture-to-path to our predictive Markov models offers the possibility of going beyond the limits of experimental observations and offering accurate \emph{in silico} trajectories for studying the navigational role of distinct coarse-grained behaviors. To do so, however, we must first connect posture deformations with movement in the environment.

Following previous work \cite{Keaveny2017}, we approximate the interaction between the worm's body and the viscous agar surface through resistive force theory (RFT) \cite{Gray1955}. This phenomenological approach assumes that each segment along the body experiences independent drag forces. Despite its simplicity, this approximation has been successfully applied to predict the motility of various organisms in viscous fluids \cite{Lauga2006,Liu2011,Bayly2011} and granular materials \cite{Zhang2014}.

\begin{figure*}[ht!]
\begin{center}
\includegraphics[scale=1]{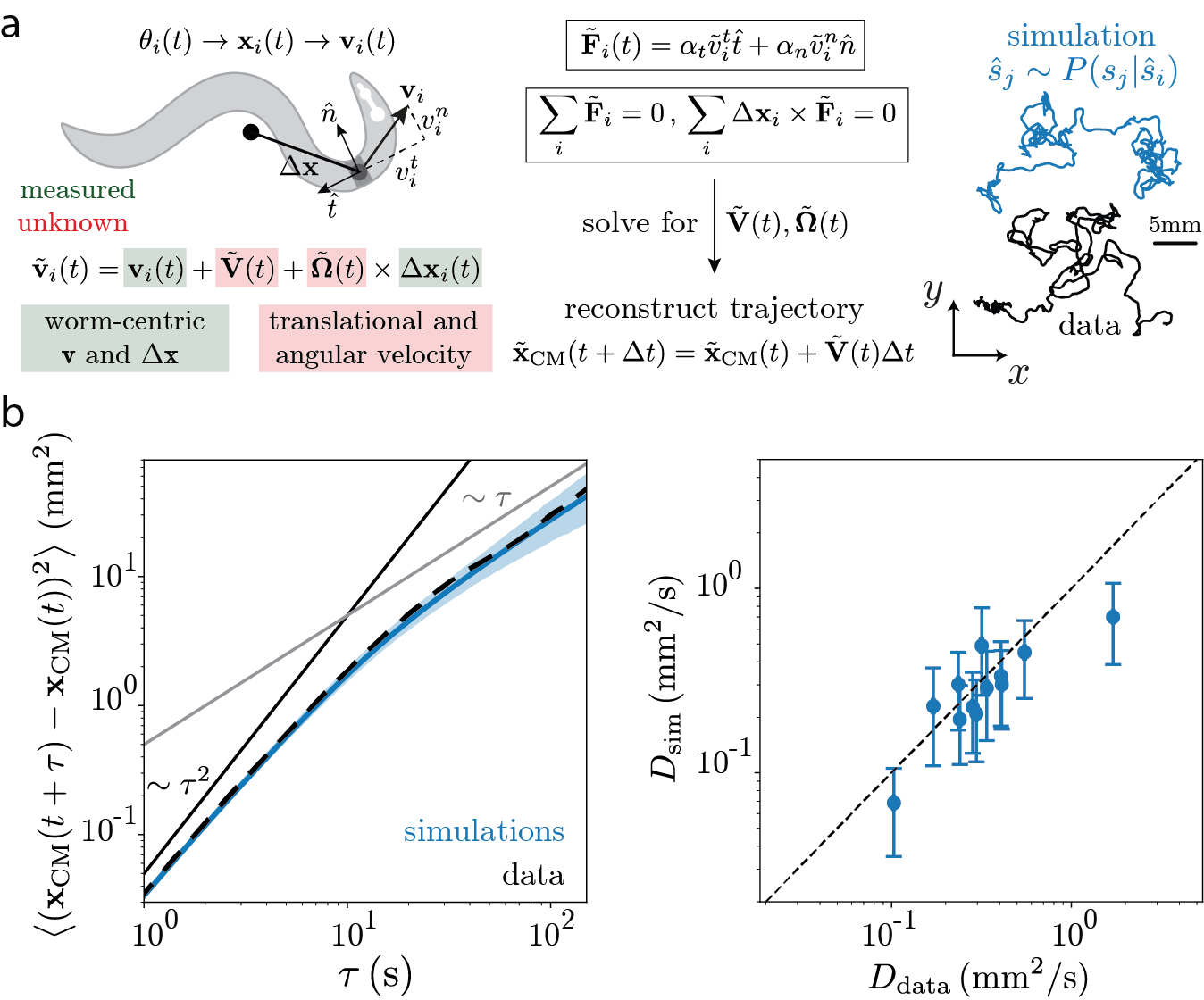}
\caption{{\bf Recovering foraging trajectories from posture simulations. }
(a) We use resistive force theory (RFT) \cite{Keaveny2017} to translate the simulated posture dynamics into movement. We simulate the Markov chain posture dynamics as described in Fig.~\ref{fig_2}. At each frame $t$ of the simulated posture time series, we reconstruct the coordinates of the $i$-th segment of the skeleton, ${\bf x}_i(t)$, from the tangent angles $\theta_i(t)$, which are themselves a linear combination of eigenworms with the mode weights particular for each frame. The measured velocities, ${\bf v}_i$, in the frame-of-reference of the worm, correspond to subtracting the overall velocity $\tilde{\bf V}(t)$ and angular velocity $\tilde{\bf \Omega}(t)$ of the worm from the lab-frame velocities $\tilde{\bf v}_i(t)$, which are unknown. Following the results Ref. \cite{Keaveny2017}, we recover the lab-frame translational $\tilde{\bf V}(t)$ and angular  $\tilde{\bf \Omega}(t)$ velocities by leveraging resistive force theory to equate the tangent and normal forces acting in each local segment to the local velocities and imposing a zero net-force and net-torque condition. The ratio between tangent $\alpha_t$ and normal $\alpha_n$ damping coefficients is the single free parameter $\alpha = \alpha_t/\alpha_n$, which we find by minimizing the distance between simulated and real trajectories, Fig.\,S5(a). We show an example worm trajectory (black), as well as simulated trajectories reconstructed from posture time series generated from the operator dynamics (blue). 
(b-left) Mean square displacement of centroid trajectories obtained from model simulations (blue) and data (black) for an example worm. Simulation error bars are 95\% confidence intervals of the mean bootstrapped across 1000 simulations. The results for all 12 worms analyzed here can be found in Fig.\,S5(b).
(b-right) Effective diffusion coefficients obtained from simulations and from the data. We estimate $D$ by fitting the slope of the mean square displacement curves in the range $\tau\in[60,100]\,\text{s}$, $\text{MSD}(t) = 4Dt$. Errors are standard deviations of the estimated diffusion coefficients across simulations.
}
\label{fig_3}
\end{center}
\end{figure*}

To propel the worm, we first reconstruct the skeleton positions in each frame ${\bf x}_i(t)$ from the instantaneous tangent angles $\theta_i(t)$ along each body segment $i$, Fig.\,\ref{fig_3}(a-left). From these we derive the worm-centric velocities ${\bf v}_i(t) = {\bf x}_i(t+1) - {\bf x}_i(t)$ and displacements with respect to the center-of-mass position ${\bf x}_\text{CM}$, $\Delta{\bf x}_i(t) = {\bf x}_i(t) - {\bf x}_\text{CM}(t)$. This results in an expression for the underlying velocities at each body segment as a function of the measured worm-centric ${\bf v}$ and $\Delta {\bf x}$ and unknown overall translational $\tilde{\bf V}(t)$ and angular $\tilde{\bf \Omega}(t)$ velocities. As in \cite{Keaveny2017}, we use linear resistive force theory to decompose the force acting on each body segment into tangent and normal components $\tilde{\bf F}_i(t) = \alpha_t \tilde{v}_i^t \hat{t}+\alpha_n \tilde{v}_i^n \hat{n}$, with drag coefficients $\alpha_n$ and $\alpha_n$, Fig.\,\ref{fig_3}(a-middle). Using this approximation, we can then recover the unknown underlying velocities $\tilde {\bf v}_i(t)$ by imposing a zero net force and torque condition. The only free parameter in this model is the ratio between the normal and tangential drag coefficients $\alpha = \alpha_n/\alpha_t$, which we infer by minimizing the distance between the reconstructed centroid trajectories and the real data (see SI Appendix: Methods), Fig.\,S5(a). In agreement with the results of Keaveny et al. \cite{Keaveny2017}, we find that in such food-free conditions, the value of $\alpha$ that optimizes the reconstruction of centroid trajectories is $\alpha^* = 30\,(29,31)$. Using such $\alpha^*$, we then reconstruct the centroid path corresponding to the posture time series simulated according to our Markov chain, and show that these qualitatively resemble real worm trajectories, Fig.\,\ref{fig_3}(a-right).

To further quantify the similarity between the centroid trajectories reconstructed from posture simulations and the data, we estimate the mean squared displacement $\text{MSD}(\tau) = \langle({\bf x}_\text{CM}(t+\tau)-{\bf x}_\text{CM}(t))^2\rangle$, which exhibits a transition between super-diffusive (nearly ballistic) and diffusive behavior between 10\,s and 100\,s \cite{Helms2019,Stephens2010,Salvador2014,Alves2017}, Fig.\,\ref{fig_3}(b-left), Fig\,S5(b). The foraging trajectories corresponding to the operator-based simulations accurately capture the MSD across a wide range of scales, including the ballistic-to-diffusive transition. To further assess the quality of the simulations, we estimate an effective diffusion coefficient by fitting $\text{MSD} = 4D\tau$ in the linear regime\footnote{We note that on longer time scales the MSD exhibits the behavior of a confined random walk due to the rigid boundaries of the agar plate, which makes it non-trivial to accurately estimate the diffusion coefficient \cite{Artale1997}. We fit the diffusion coefficient in the regime $\tau \in [60,100]$, which corresponds to a time scale within which finite-size effects are negligible and the mean squared displacement is approximately a linear function of $\tau$, $\text{MSD} \sim \tau$.} and find that, across worms, the resulting diffusion coefficients obtained from simulations closely follow the data, Fig.\,\ref{fig_3}(b-right). Our results demonstrate that it is possible to predict long time scale diffusive properties from fast posture dynamics in animals.

\section*{Coarse-graining behavior through ensemble dynamics}

As highlighted in Fig.\,\ref{fig_2}, $\emph{C. elegans}$ foraging behavior exhibits multiple time scales: from the body waves that define short-time behaviors (e.g., forward, reversal, turns), to longer-time sequences (e.g., run, pirouette) that the worm uses to navigate its environment. Typically, these longer-time sequences have been identified phenomenologically by setting thresholds on heuristically defined quantities,  as was done in Fig.\,\ref{fig_2}(c). Here, we show that it is possible to reveal the multiple scales of \emph{C. elegans} locomotion directly from the posture dynamics. Intuitively, stereotyped behaviors correspond to regions of the behavioral space that the animal visits often. In an analogy with statistical mechanics, we can imagine behavior as evolving on a complex potential landscape, where each well corresponds to a particular stereotyped behavior, and the barrier heights set the transition time scales. Such a picture emerges naturally when analyzing the dynamics through an ensemble approach, and we leverage our inferred Markov chain to directly identify metastable behaviors.

\subsection*{``Run-and-Pirouette''}

\begin{figure*}[ht!]
\begin{center}
\includegraphics[scale=1]{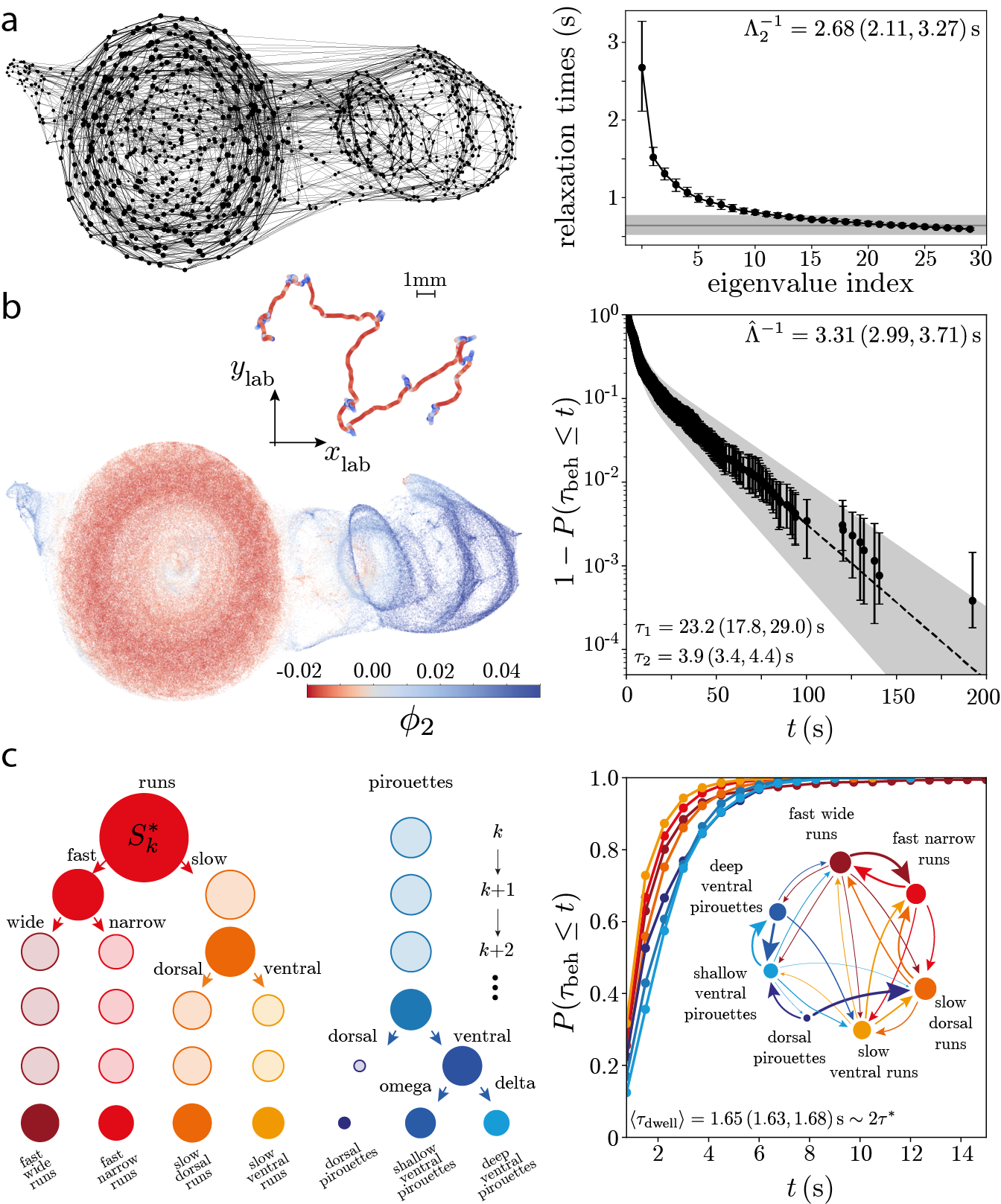}
\caption{{\bf Slow modes for coarse-graining: from ``runs-and-pirouettes'' to stereotyped body waves.}
(a-left) Schematic of the inferred Markov chain. We partition the maximally predictive sequence space $X_{K^*}$ ($11\times 5$ dimensions) into Voronoi cells, here represented as points in the 2-dimensional UMAP embedding space. The size of each point is proportional to the probability of visiting a given microstate $s_i$, and the line width corresponds to the probability of transitioning among distinct states after $\tau=1\,\text{frame}$ (we only show transitions with $P_{ij}>0.025$ for simplicity).  (a-right) Relaxation timescales obtained from the 30 eigenvalues with the largest real part of $P_{ij}(\tau^*)$ with $\tau^* = 0.75\,\text{s}$. The horizontal gray bar is the largest non-trivial eigenvalue of the transition matrix computed from a shuffled symbolic sequence (see SI Appendix: Methods for details). Error bars are 95\% confidence intervals bootstrapped across worms. 
(b-left) We color the sequence space and an exemplar centroid trajectory (inset) by the first nontrivial eigenvector of the reversibilized transition matrix $\phi_2$, which optimally separates metastable states \cite{Froyland2005,Costa2023}. Positive values correspond to combinations of reversals and turns that reorient the worm, these are ``pirouettes'', while negative values correspond to ``forward runs''.
(b-right) We use $\phi_2$ to identify two coarse-grained regions, which we denote as ``macroscopic'' behavioral states, and find that the complementary cumulative distribution function of the resulting dwell times $1-P(\tau_\text{dwell}\leq t)$ is characterized by two time scales, which we extract by fitting a sum of exponential functions. The inferred time scales are in agreement with previous phenomenological observations of worm behavior \cite{Pierce-Shimomura1999}, and result in a relaxation time $\hat{\Lambda}^{-1} = (\tau_1^{-1}+\tau_2^{-1})^{-1} = 3.31\,(2.99,3.71)\,\text{s}$ \cite{VanKampen1992} that agrees with the largest eigenvalue of $P_{ij}(\tau^*)$, $\Lambda_2^{-1}$, within error bars. The timescale errors and error bars are 95\% confidence intervals bootstrapped across events.
(c) The subdivision process. At each iteration step, we subdivide the metastable state with the largest measure, $S_k^*$, along the first nontrivial eigenvector obtained from the transition probability matrix conditioned only on the states within that metastable state \cite{Ma2013}. This results in a top-down subdivision of behavior that follows the structure of an effective free energy landscape. The size of the circles represents the relative measure in each state. For interpretability, we stop at the 5th subdivision, yielding 7 ``mesoscopic'' states (characterized below). (c-right) Cumulative distribution of the mesoscopic state dwell times: the duration is short $(\approx 2\tau^*)$. The inset shows a transition diagram in which the size of the nodes and the edge widths are proportional to the measure in each behavior and the transition probabilities, respectively (transition probabilities $<0.05$ are not graphed for simplicity). 
}
\label{fig_4}
\end{center}
\end{figure*}

The eigenvalues of the transition matrix provide direct access to the long time-scale properties of the dynamics, even when these are not directly apparent from the original trajectories or the equations of motion. The real part of the eigenvalues $\{\lambda_i\}$ of $P_{ij}(\tau^*)$ quantify the relaxation time to the steady state,
\begin{align}
\Lambda_i^{-1} = \frac{-\tau^*}{\log \operatorname{Re}(\lambda_{i})}.
\label{eq:t_imp}
\end{align}
The spectrum of relaxation times is shown in Fig.\,\ref{fig_4}(a-right), and exhibits both an isolated, longer-lived mode with $\Lambda_2^{-1} = 2.68\,(2.11,3.27)\,\text{s}$, as well as an additional $\approx10$ significant modes.

The eigenvectors corresponding to the long-lived dynamics provide continuous reaction coordinates for a natural coarse-graining. In Fig.~\ref{fig_4}(b) we color-code the maximally-predictive sequence space according to $\phi_2$; 
comparing with Fig.~\ref{fig_1}(c), we find that negative values of $\phi_2$ (red) correspond to positive phase velocities and low curvatures, indicative of ``forward runs'', while positive values of $\phi_2$ (blue) generally correspond to sequences of reversals, dorsal and ventral turns used during abrupt reorientations, see inset of Fig.~\ref{fig_4}(b). The inset shows an example 10 minute long centroid trajectory color coded by the projection along $\phi_2$. Negative projections occur during ``runs'', while positive values are found during abrupt reorientation events composed of sequences of reversals and turns. We thus obtain a slow reaction coordinate that captures the dynamics along a posture-based expression of a ``run-and-pirouette'' axis. 

The remaining eigenfunctions reveal substructures within ``runs'' and ``pirouettes''. For example, in Fig.\,S6 we show the maximally-predictive sequence space color coded by the following 3 long-lived eigenvectors, $\{\phi_3,\phi_4,\phi_5\}$. We see that $\phi_3$ differentiates dorsal and ventral turns, $\phi_4$ differentiates turning and reversals, and $\phi_5$ differentiates the compound motifs of shallow turns following a pause, from reversals that are followed by deep $\delta$-turns. Projections onto the slow eigenvectors thus reveal continuous modulations along the short nematode locomotory movements first observed by Croll \cite{Croll1975}, Fig.\,S6, as well as longer-lived, compound motifs, such as runs-and-pirouettes, Fig.\,\ref{fig_4}(b). 

While the continuous variation along these slow modes neatly highlights the temporal organization of distinct features of worm behavior, we can also use them to identify stereotyped behavioral states as ``macroscopic'' metastable sets \cite{Schutte2001,Bollt2013}: groups of microstates that transition more often within rather than between groups. Stereotyped behaviors correspond to finescale movements that occur more often than not in a sequence. In other words, stereotypy implies a timescale separation between variations on what we call a behavioral state, and transitions among distinct stereotyped behaviors. This notion of stereotypy is naturally captured by the notion of coherent/metastable sets, which can be directly identified through the eigenvalues and eigenvectors of the Markov chain. In this way, the structure of these sets and the kinetics between them offer a coarse-graining of behavior that directly follows the multiple timescales of the dynamics. We note that this construction provides a precise meaning to the discreteness of such behavioral states: the discrete approximation is appropriate at a particular timescale of interest. Faster timescales require finer scale states, while longer timescales can be neatly approximated by larger coarse-grained states.

\begin{figure*}[ht!]
\begin{center}
\includegraphics[scale=1]{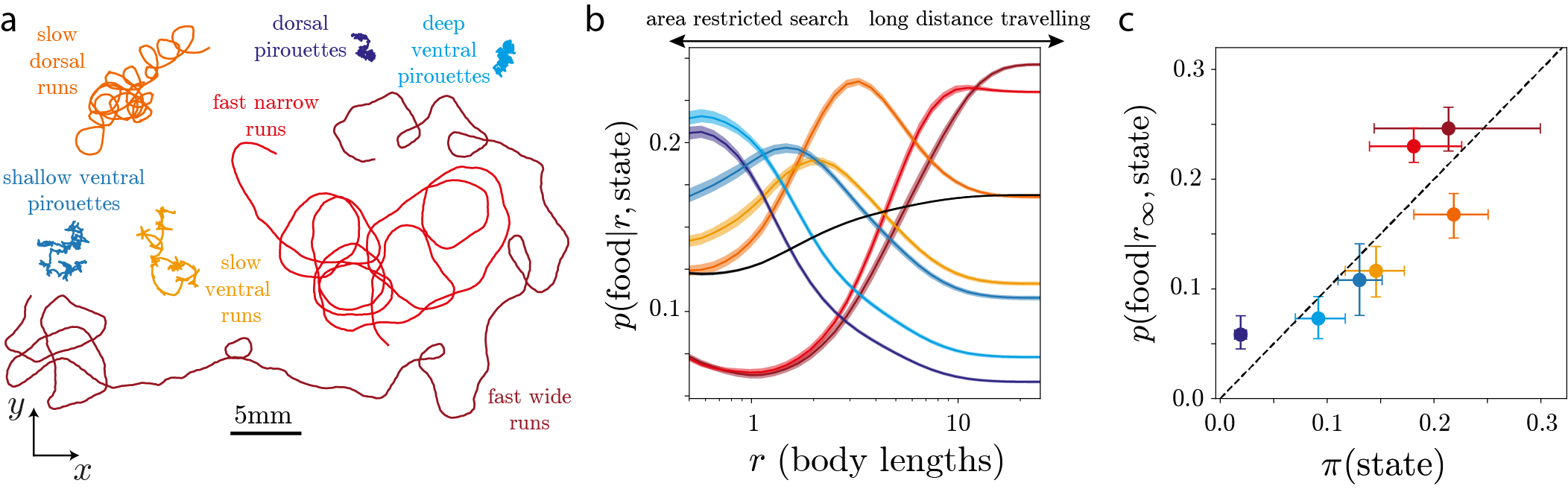}
\caption{{\bf Leveraging our multiscale description to explore the role of behavioral states. }
We simulate long trajectories while forcing the worm's dynamics to remain within a single mesoscopic state $S$ obtained through a subdivision of the behavioral space into 7 states, Fig.\,\ref{fig_4}(c). We proceed as in Figs.\,\ref{fig_2},\ref{fig_3}, but sample each new state according to a transition matrix built only within the partitions belonging to $S$, $\hat{s}_j(t+\tau) \sim P_S(s_j|\hat{s}_i(t)),\, i,j\in S$.
(a) Example 10 min long centroid trajectories of each state. 
(b) Probability of finding food in a uniform food patch of radius $r$ for a particular state. We find that the most efficient behavioral states change depending on the radius of the food distribution: at short distances, ``pirouette'' and slow ``run'' states provide better chances of finding food, whereas at larger distances the two fast ``run'' states are most efficient. The black line represents the strategy of an average worm $p(\text{food}|r,\text{average worm}) = \sum_S \pi(S) p(\text{food}|r,S)$, obtained as a weighted average over the probability of finding a worm in a particular mesoscopic state $S$, $\pi(S)$. While the worm's movement strategy is more likely to find food sources scattered at larger distances, the ensemble of mescoscopic states can be utilized for foraging success across scales.
(c) The likelihood of finding food in a large uniform food patch for each of the states closely matches the probability of finding a worm in a particular state.
}
\label{fig_5}
\end{center}
\end{figure*}
More principedly, we start by identifying long-lived behavioral states as almost invariant sets \cite{Froyland2005} in the reconstructed state space.  These sets are optimally determined by the second eigenvector $\phi_2$ of the time-symmetric reversibilized transition matrix $P_r$, which we infer from the ensemble of worms (see SI Appendix: Methods) \cite{Dellnitz1999,Froyland2003}. 
We search along $\phi_2$ for a single threshold that maximizes the metastability of both resulting coarse-grained sets (see SI Appendix: Methods) \cite{Costa2023}, Fig.\,S7, and we identify the resulting two macrostates as ``run'' and ``pirouette''.  The complementary cumulative distribution of the time spent in either of the two states $1-P(t_\text{beh}\leq t)$ is roughly characterized by two time scales, Fig.\,\ref{fig_4}(b-right), fit by a sum of exponential functions and in excellent agreement with previous phenomenological observations \cite{Pierce-Shimomura1999}. In addition, these transition timescales are related to the timescale of relaxation to the steady state distribution as $\hat{\Lambda}^{-1} = 1/(\tau_1^{-1}+\tau_2^{-1}) = 3.313\,(2.985,3.709)\,\text{s}$ \cite{VanKampen1992}, which agrees with the relaxation times of the transition matrix within statistical accuracy, Fig.\,\ref{fig_4}(a-right). In our analysis ``run-and-piroutte'' kinetics emerge directly from worm-centric posture dynamics, without any positional information.

\subsection*{ ``Run(s)-and-Pirouette(s)''}

While dividing the dynamics along $\phi_2$ identifies the longest lived states, the existence of other significant eigenvectors, Fig.\,\ref{fig_4}(a), indicates that there are finer-scale divisions.  To identify such states in a way that naturally reflects the behavioral landscape, we iteratively subdivide the largest metastable state into faster mesoscopic states \cite{Ma2013}, Fig.\,\ref{fig_4}(c). In practice, we identify the most-visited metastable state, construct a reversibilized transition matrix using only the microstates within that metastable set, and use its first nontrivial eigenvector to subdivide the dynamics (see SI Appendix: Methods for details). This is akin to subdividing a free-energy landscape; at each iteration, we subdivide the system along the largest energy barrier within the highest measure basin. We also note that our subdivisioning proceeds {\em from the longest-lived states down} rather than from the shortest-lived states up, where the latter is more common in behavior coarse-graining approaches \cite{Brown2013,Berman2014a,Reddy2022}.

In the foraging behavior of {\em C. elegans}, beyond the initial division into runs and pirouettes (which we denote as ``macroscopic'' states), we further subdivide the dynamics into 7 ``mesoscopic'' interpretable states: 4 distinct run states and 3 subdivisions of the pirouette state Fig.\,\ref{fig_4}(c),\,S8. The run state essentially splits into two fast states and two slower states, which can be distinguished either by the wave length of the body, or by having a particular bias towards the dorsal or ventral sides: the dorsally-biased slow state is akin to a head-casting state \cite{KAPLAN2020}, while the ventrally-biased stated is akin to a ``dwelling'' state \cite{Fujiwara2002,BenArous2009,Flavell2013}, with incoherent head and tail movements and no propagating wave \cite{Hebert2021}. On the other hand, the pirouette state neatly splits into dorsal turns, deep ventral $\delta$-turns, and reversals followed by shallow $\Omega$-turns.  

These mesoscopic states that decorate the worm's foraging behavioral landscape are short-lived, with a characteristic time scale of $\langle \tau_\text{dwell} \rangle = 1.65\,(1.63,1.68)\,\text{s} \approx 2\tau^*$, Fig.\,\ref{fig_4}(c-right). The transition diagram between them, Fig.\,\ref{fig_4}(c-right,\,inset), reveals the fine-scale organization of the worms' foraging strategy. Further subdivisions result in even shorter-lived states, which are increasingly challenging to interpret.

\section*{Exploring the role of the mesoscopic states }

In the data analyzed here, worms were grown in a food-rich environment, but then placed on food-free agar plates and allowed to move without restrictions. Under these conditions, the worm's behavior has been qualitatively described as foraging \cite{Hills2004,Calhoun2014}. We apply our approach to better understand the role of the mesoscopic states in the worm's search for food.

We use the Markov model to simulate \emph{in silico} worms that are forced to remain in a particular mesoscopic state and the posture-to-path framework to investigate the properties of the trajectories resulting from the posture dynamics in each of the states. We can simulate trajectories that are much longer than those observed in the data, Fig.\,\ref{fig_4}(c-right), allowing us to dissect how different states produce distinct large-scale tracks. In Fig.~\ref{fig_5}(a), we show simulated $10\,\text{min}$ long trajectories for each of the 7 mesoscopic states. Notably, the difference in posture wavelengths exhibited by the two fast ``run'' states, Fig.\,S8(b), results in dramatically different trajectories, with the longer wavelength state (fast wide runs) resulting in overall straighter paths, and the shorter wavelength state (fast narrow runs) resulting in ventrally-biased curved trajectories with a diameter that is several times the body length and a period orders of magnitude longer than the body wave period, Fig.\,S9. Interestingly, the dorsally-biased slow state also results in loopy trajectories, but with a shorter diameter and faster recurrence time. In addition, the ventrally-biased ``dwelling''-like slow state \cite{Fujiwara2002,BenArous2009,Flavell2013} with its frequent head retractions results in a denser sampling of a local patch. Finally, the three ``pirouette'' states result in a denser sampling of space and a reduced centroid displacement.

We next interrogate the efficiency of each of the 7 mesoscopic states at encountering food uniformly distributed within a disc of radius $r$ around the initial position, Fig.\,\ref{fig_5}(b), a simple but informative condition. We find that the ``pirouette'' states as well as the slow ``run'' states are most efficient at finding food at shorter distances, while at larger distances the two fast ``run'' states perform best. Such a differential use of behaviors, evocative of an exploration-exploitation trade-off, is also seen in nature. Upon encountering food, \emph{C. elegans}, as well as many other species, engage in area restricted search, which is characterized by shorter paths and a high frequency of large angle turns \cite{Kareiva1987,Bell1990,Hills2004,Gray2005,Salvador2014,Calhoun2015,Hums2016,Flavell2020}. Conversely, upon removal from food, \emph{C. elegans} lowers its turning rate \cite{Gray2004,Stephens2010} to engage in global search or long distance travel \cite{Hills2004,Gray2005,Calhoun2015,Hums2016,Flavell2020}.

Remarkably, we find that instead of only using the most efficient behavioral state (``fast wide runs''), worms engage in a strategy that employs each mesoscopic state in a proportion that closely matches the relative efficiency of the different states at finding food uniformly distributed in a large patch (several body lengths), Fig.\,\ref{fig_5}(c). This ``probability matching'' behavior has been observed across several species, including humans (see, e.g., \cite{Neimark1959,Bullock1962,Bitterman1958,Vulkan2000,Newell2017,Saldana2022}), and emerges naturally in ``multi-armed bandit'' situations in which agents must decide among different actions that yield variable amounts of reward without knowing \emph{a priori} the relative reward of each action (see, e.g., \cite{Slivkins2019}).

\section*{Discussion}

We combine maximally-predictive short posture sequences with a Markov chain model to bridge disparate scales in the foraging dynamics of the nematode worm {\em C. elegans}. Rather than seeking low-dimensional descriptions of the data directly (e.g.~\cite{Flavell2013,Schwarz2015,Gomez-Marin2016,Salvador2014,Berman2016,Gupta2019}), we instead first {\em expand} in representation complexity: enlarging the variable of interest to include time in the form of posture {\em sequences} and constructing a maximum entropy partition to capture as much predictive information as possible. This expansion both in time and number of microstates is similar in spirit to that currently found in large language models, though our conceptual approach is dramatically simpler. 

The maximally-predictive sequence space combines worm postures from roughly a quarter of the duration of a typical body wave, in agreement with previous work \cite{Ahamed2021}.  On longer timescales, the posture-based ``run-and-pirouette'' navigation strategy \cite{BERG1972,Pierce-Shimomura1999} derived from the inferred Markov dynamics provide an accurate and principled coarse graining of foraging behavior, disentangling motions that are confounded by centroid-derived measurements (see e.g.~\cite{Roberts2016}). This is particularly evident in our subdivision of the behavioral space. For example, we identify distinct ``run'' gaits that exhibit comparable centroid speeds, but are clearly distinguishable by the posture dynamics. Additionally, our top-down subdivision of behavior reflects the hierarchy of timescales in \emph{C. elegans} foraging behavior \cite{KAPLAN2020}. Our approach systematically identifies such a control hierarchy from behavioral recordings alone, connecting  posture timescales to ``run-and-pirouette'' kinetics. It will be interesting to investigate how the mesoscopic states identified here are controlled by the nervous system of the worm, and recent advances in experimental techniques that permit simultaneous neural and behavioral imaging in \emph{C. elegans} provide an exciting path toward such discoveries \cite{Kato2015,Venkatachalam2016,Hallinen2021,Yemini2021,Atanas2023}. 

The power of our modeling approach is in its simplicity; we bridge scales using a simple but effective Markov model, and this is only possible by recognizing and exploiting the mutual dependence between modeling and representation. Instead of directly modeling the posture time series (which can require higher-order and highly non-linear terms, see e.g.~\cite{Stephens2008}), we search for maximally predictive states such that a simpler Markovian description can nevertheless accurately predict behavior. These emergent Markov dynamics offer a promising and powerful demonstration of quantitative connections across the hierarchy of movement behavior generally exhibited by all organisms \cite{Berman2016,Tinbergen1963}. 

By finely partitioning the space of posture sequences, we encode continuous nonlinear dynamics through a Markov chain with a large number of states. This is analogous to building a hidden Markov model (HMM), but one in which the ``hidden'' states are actually observable (through time delays of our observations), and for which there is a one-to-one correspondence between ``hidden'' states and emitted symbols: each observation in the posture sequence $\vec{a}(t)$ uniquely determines the state $X_{K^*}(t)$.  While HMMs are commonly used in behavioral analysis (see e.g. \cite{Gallagher2013,Hebert2021}), they are rarely built with so many states and with the goal of correctly predicting dynamics. In particular, most approaches employ a small number of discrete behavioral states, where the number of states is a hyperparameter of the model and the discretization is not unique. In contrast, we let the data \emph{reveal} the ``hidden'' states through time delays, and set the discretization so as to maximize predictive information. In this sense, the HMM we build is unique: by revealing the \emph{hidden} dynamics through time delays, the ``hidden'' states are uniquely determined by the observations, making the HMM unifilar \cite{Jurgens2021}.  In other words, the ``hidden'' states themselves have a very definite meaning in our approach: we effectively group together ``pasts'' that have equal predictability over the future up to an $\epsilon$-resolution (set by the number of partitions $N\sim \epsilon^{-D_\text{emb}}$, where $D_\text{emb}$ is the intrinsic embedding dimension of the dynamics), approximating the system's \emph{causal states} \cite{Shalizi2001}. This set of states together with the resulting Markov chain effectively constitutes an $\epsilon$-machine \cite{Crutchfield1989}, the minimal maximally-predictive machine. Any other HMM in which hidden states are not \emph{causal} returns models that severely overestimate the complexity of the dynamics. In addition, even though we start with a large transition matrix, we can coarse-grain it by identifying which states commonly follow each other in time to generate stereotyped sequences. In this way, instead of imposing discrete states from the start (as is common with HMM approaches), we first identify a large number of predictive causal states and only then leverage the resulting Markovian dynamics to identify coarse-grained stereotyped behaviors. 

Our information theoretic framework also frees us from the constraint of linearity that is commonly imposed in graphical models applied to animal behavior (such as autoregressive hidden Markov models \cite{Wiltschko2015}). In particular, while the stereotyped states found through such models are encoded by linear dynamics, the states we identify can exhibit much more complex nonlinear dynamics, allowing us to capture longer time-scale structures in behavior. 

Are Markov models enough to capture the richness of animal behavior more universally?  It is important to distinguish between two sources of non-Markovianity. The first one is general and is simply induced by the fact that time series data are typically only a partial observation of the full dynamical state \cite{Costa2023,Rupe2022}. Projecting the full unobserved dynamics onto a subset of observable degrees of freedom inherently results in non-Markovian dynamics for the measured variables \cite{Mori1965,Zwanzig1973,Lin2021,Gilani2021}. Such an ``under-embedding'' might result in apparent memory when naively constructing behavioral states. The second source of non-Markovianity, which is less trivial and likely ubiquitous in behavior, derives from the fact that there may be ``hidden'' latent variables that modulate behavior over timescales comparable to the measurement time \cite{Bialek2023_PRL}. In this case, the steady-state distribution itself is changing slowly over time, rendering the dynamics explicitly non-stationary \cite{costa2023fluctuating}.

A relevant demonstration of non-stationarity is provided by the adaptive changes in pirouette rate seen in the behavior of {\em C. elegans} upon removal from a food-rich environment \cite{Hills2004,Gray2005,Stephens2010, Salvador2014}. This adaptation is present in the data we analyze and is not captured by our Markov model \cite{costa2023fluctuating}, Fig.\,S10. To characterize such non-ergodic latent variables requires explicitly time-dependent Markov models, which we leave for future work. We note, however, that our coarse graining can be easily extended to capture non-stationary dynamics through the discovery of $\tau$-dependent coherent sets that identify \emph{moving} regions of the state space that remain coherent within a time scale $\tau$ \cite{Froyland2010,Froyland2010b,Wang2015,Koltai2016,Koltai2018,Fackeldey2019}.

Particularly interesting future directions include the analysis of even longer dynamics in {\em C. elegans} \cite{Hebert2021,Martineau2020,Ikeda2020, Le2020}, where we expect to be able to extract longer-lived behavior strategies, such as the minutes-long transitions between ``roaming'' and ``dwelling'' states in food-rich environments \cite{Fujiwara2002,BenArous2009,Flavell2013}. Our modeling approach can also be used as means to obtain a deeper understanding of the effects of genetic, neural, or environmental perturbations on the multiple scales of \emph{C. elegans} behavior. Indeed, the inferred transition matrices are a powerful phenotype that encapsulates multiple scales of \emph{C. elegans} foraging behavior, and has the power to reveal how behavior is affected by a given perturbation. Of particular relevance for the study of long timescales in behavior would be to focus on mutations that impair neuromodulatory pathways and are thus likely to impact the spectrum of relaxation times of the inferred Markov chain.

The effectiveness of our Markov model at capturing the nonlinear dynamics of \emph{C. elegans} body pose, combined with the ability to translate those spatiotemporal dynamics into movement, have allowed us to investigate how different behavioral states result in distinct ways of exploring the environment at much larger scales. Our analysis recovered the two main foraging modes exhibited by \emph{C. elegans}: one that combines different pirouette states and slow runs resulting in a local search, and another one that mostly leverages fast run states to search for food more globally \cite{Hills2004,Gray2005,Salvador2014,Calhoun2015}. 

We also discovered that the relative use of different behavioral states closely follows the relative efficiency of each state in food discovery. In fact, instead of using the behavioral state that would maximize its chances of finding food in its environment, worms match their strategy with the relative efficiency of each state. Interestingly, such a strategy, termed probability matching \cite{Newell2017}, or Thompson sampling \cite{Thompson1933}, is a well-studied heuristic solution for the multi-armed bandit problem, a game in which different actions have variable rewards that are \emph{a priori} unknown to the player, whose goal is to maximize total pay-out. Evidence for probability matching in decision making tasks has been previously demonstrated in experiments in animals and humans \cite{MaBouDi2020,Beron2022}, and is an active area of research in cognitive science of decision making \cite{Schulz2020}. 

While this strategy seems ``irrational'' in the context of maximizing reward in a fixed environment, that is not the condition in which worms have evolved: in ecologically-relevant situations the environment changes over time, rendering the distribution of rewards non-stationary and the subsequent sampling events correlated. Interestingly, reinforcement learning agents that have been evolved in a changing environment also develop probability matching strategies \cite{Niv2002}. In addition, it has been shown that optimal Bayesian learners engage in probability matching when they expect sampling events to have temporal dependencies \cite{Green2010}, as is the case in most ecologically relevant scenarios in which samples of the environment are not independent, but exhibit temporal correlations (at least as a result of the actions taken by the agent). This suggests that probability matching may reflect an exploration-exploitation trade-off that robustly maximizes reward in an ever-changing environment. Our results indicate that \emph{C. elegans} may implement such a heuristic in its foraging strategy. If the worm is indeed probability matching, it may have a way of storing its estimate of the current probability of success for each strategy (which may be reflected the dynamics itself).  It will be fascinating to look for signatures of this in situations where we can experimentally adjust the pay-out probabilities.  It may also be possible in a tractable organism like \emph{C. elegans} to estimate metabolic costs to utilize each behavioral state \cite{Braeckman2002}.  The multitude of genetic tools \cite{Boulin2012}, the ability to image neurons in behaving animals \cite{Kato2015,Venkatachalam2016,Hallinen2021,Yemini2021,Atanas2023} and to quantify behavior in detail makes \emph{ C. elegans} an ideal system to look inside of an organism as it experiences the world, makes decisions based on those observations and its internal model, and updates that internal model based on the outcomes of those decisions and their effect on the environment.

\section*{Materials and Methods}

Details of the reconstruction of maximally-predictive posture sequences, Fig.\,\ref{fig_1}, and the top-down subdivision of \emph{C. elegans} behavior, Fig.\,\ref{fig_4}, can be found in SI Appendix, and build upon the implementation of \cite{Costa2023}. The SI Appendix also includes additional information about the simulation of symbolic sequences and posture time series presented in Fig.\,\ref{fig_2}. Finally, the details of the posture-to-path simulations used in Figs.\,\ref{fig_3},\,\ref{fig_5} can also be found in the SI Appendix and build upon the implementation of \cite{Keaveny2017}.

\noindent {\bf Software and data availability:} Code for reproducing our results is publicly available:
\url{https://github.com/AntonioCCosta/markov_worm/}. Data can be found in \cite{manuscript_data_doi}.

\section*{Acknowledgements}
We thank Massimo Vergassola and Federica Ferretti for comments. This work was supported by OIST Graduate University (TA, GJS), a program grant from the Netherlands Organization for Scientific Research (AC, GJS), by the Herchel Smith Fund (DJ), and by Vrije Universiteit Amsterdam (AC, GJS). GJS acknowledges useful (in-person!) discussions at the Aspen Center for Physics, which is supported by National Science Foundation Grant PHY-1607611.

\bibliographystyle{pnas-new}
\bibliography{Bibliography}

\begin{thebibliography}{100}

\bibitem{Weinert2016}
Weinert F (2016) {\em Laplace's Demon: Causal and Predictive Determinism}.
\newblock (Springer International Publishing, Cham), pp. 65--72.

\bibitem{Mori1965}
Mori H (1965) {Transport, Collective Motion, and Brownian Motion}.
\newblock {\em Progress of Theoretical Physics} 33(3).

\bibitem{Zwanzig1973}
Zwanzig R (1973) {Nonlinear generalized Langevin equations}.
\newblock {\em Journal of Statistical Physics} 9(3):215--220.

\bibitem{Berman2018}
Berman GJ (2018) {Measuring behavior across scales}.
\newblock {\em BMC Biol.} 16(23).

\bibitem{Bialek2023_PRL}
Bialek W, Shaevitz JW (2024) Long timescales, individual differences, and scale
  invariance in animal behavior.
\newblock {\em Phys. Rev. Lett.} 132(4):048401.

\bibitem{Costa2023}
Costa AC, Ahamed T, Jordan D, Stephens GJ (2023) {Maximally predictive states:
  From partial observations to long timescales}.
\newblock {\em Chaos: An Interdisciplinary Journal of Nonlinear Science}
  33(2):023136.

\bibitem{Rupe2022}
Rupe A, Vesselinov VV, Crutchfield JP (2022) Nonequilibrium statistical
  mechanics and optimal prediction of partially-observed complex systems.
\newblock {\em New Journal of Physics} 24(10):103033.

\bibitem{Brenner1974}
Brenner S (1974) {The Genetics of Caenorhabditis elegans}.
\newblock {\em Genetics} 77(1):71--94.

\bibitem{Bargmann1993}
Bargmann CI (1993) Genetic and cellular analysis of behavior in c. elegans.
\newblock {\em Annual Review of Neuroscience} 16(1):47--71.
\newblock PMID: 8460900.

\bibitem{Bargmann2013}
Bargmann CI, Marder E (2013) {From the connectome to brain function}.
\newblock {\em Nat Methods} 10(6):483–490.

\bibitem{Pereira:2019te}
{Pereira, Talmo D}, et~al. (2019) {Fast animal pose estimation using deep
  neural networks}.
\newblock {\em Nature Methods} 16(1):117--125.

\bibitem{Mathis:2018us}
{Mathis, Alexander}, et~al. (2018) {DeepLabCut: markerless pose estimation of
  user-defined body parts with deep learning}.
\newblock {\em Nature Neuroscience} 21(9):1281--1289.

\bibitem{Hebert2021}
Hebert L, Ahamed T, Costa AC, O'Shaughnessy L, Stephens GJ (2021) {WormPose:
  Image synthesis and convolutional networks for pose estimation in C.
  elegans}.
\newblock {\em PLoS Computational Biology} 17(4):e1008914.

\bibitem{Pereira2020}
Pereira TD, Shaevitz JW, Murthy M (2020) Quantifying behavior to understand the
  brain.
\newblock {\em Nature Neuroscience} 23(12):1537--1549.

\bibitem{Mathis2020}
Mathis MW, Mathis A (2020) Deep learning tools for the measurement of animal
  behavior in neuroscience.
\newblock {\em Current Opinion in Neurobiology} 60:1--11.

\bibitem{Aguilar2016}
Aguilar J, et~al. (2016) {A review on locomotion robophysics: the study of
  movement at the intersection of robotics, soft matter and dynamical systems.}
\newblock {\em Reports on progress in physics. Physical Society (Great
  Britain)} 79(11):110001.

\bibitem{Brown2018}
Brown AX, de~Bivort B (2018) Ethology as a physical science.
\newblock {\em Nature Physics} 14(7):653--657.

\bibitem{Landau2014}
Landau DP, Binder K (2014) {\em A Guide to Monte Carlo Simulations in
  Statistical Physics}.
\newblock (Cambridge University Press), 4 edition.

\bibitem{Croll1975}
Croll NA (1975) Behavioural analysis of nematode movement.
\newblock {\em Adv. Parasitol.} 13:71--122.

\bibitem{Stephens2008}
Stephens GJ, Johnson-Kerner B, Bialek W, Ryu WS (2008) {Dimensionality and
  dynamics in the behavior of C. elegans.}
\newblock {\em PLoS Comput. Biol.} 4(4):e1000028.

\bibitem{Costa2019}
Costa AC, Ahamed T, Stephens GJ (2019) {Adaptive, locally linear models of
  complex dynamics}.
\newblock {\em Proceedings of the National Academy of Sciences of the United
  States of America} 116(5):1501--1510.

\bibitem{Ahamed2021}
Ahamed T, Costa AC, Stephens GJ (2021) {Capturing the continuous complexity of
  behaviour in Caenorhabditis elegans}.
\newblock {\em Nature Physics} 17(2):275--283.

\bibitem{Takens1981}
Takens F (1981) Detecting strange attractors in turbulence in {\em Dynamical
  Systems and Turbulence, Warwick 1980}, eds.{} Rand D, Young LS.
\newblock (Springer Berlin Heidelberg, Berlin, Heidelberg), pp. 366--381.

\bibitem{Sugihara1990}
Sugihara G, May RM (1990) Nonlinear forecasting as a way of distinguishing
  chaos from measurement error in time series.
\newblock {\em Nature} 344(6268):734.

\bibitem{Sauer1991}
Sauer T, Yorke JA, Casdagli M (1991) Embedology.
\newblock {\em Journal of Statistical Physics} 65(3):579--616.

\bibitem{Stark1999}
Stark J (1999) {Delay embeddings for forced systems. I. Deterministic forcing}.
\newblock {\em Journal of Nonlinear Science} 9(3):255--332.

\bibitem{Stark2003}
Stark J, Broomhead DS, Davies M, Huke J (2003) {Delay embeddings for forced
  systems. II. Stochastic forcing}.
\newblock {\em Journal of Nonlinear Science} 13(6):519--577.

\bibitem{Broekmans2016}
Broekmans OD, Rodgers JB, Ryu WS, Stephens GJ (2016) {Resolving coiled shapes
  reveals new reorientation behaviors in C. elegans}.
\newblock {\em eLife} 5(e17227).

\bibitem{Berman2014a}
Berman GJ, Choi DM, Bialek W, Shaevitz JW (2014) {Mapping the stereotyped
  behaviour of freely moving fruit flies.}
\newblock {\em J. Royal Soc. Interface} 11(99):1--21.

\bibitem{Wiltschko2015}
Wiltschko AB, et~al. (2015) {Mapping Sub-Second Structure in Mouse Behavior}.
\newblock {\em Neuron} 88(6):1121--1135.

\bibitem{McInnes2018}
McInnes L, Healy J, Melville J (2018) {UMAP: Uniform Manifold Approximation and
  Projection for Dimension Reduction, arXiv [Preprint] (2018).
  https://arxiv.org/abs/1802.03426 (accessed 19 October 2023)}.

\bibitem{Pierce-Shimomura1999}
Pierce-Shimomura JT, Morse TM, Lockery SR (1999) {The fundamental role of
  pirouettes in Caenorhabditis elegans chemotaxis.}
\newblock {\em J. Neurosci.} 19(21):9557--69.

\bibitem{Zhao2003}
Zhao B, Khare P, Feldman L, Dent JA (2003) {Reversal frequency in
  Caenorhabditis elegans represents an integrated response to the state of the
  animal and its environment}.
\newblock {\em Journal of Neuroscience} 23(12):5319--5328.

\bibitem{Gray2005}
Gray JM, Hill JJ, Bargmann CI (2005) A circuit for navigation in
  {Caenorhabditis elegans}.
\newblock {\em Proceedings of the National Academy of Sciences of the United
  States of America} 102(9):3184--3191.

\bibitem{Keaveny2017}
Keaveny EE, Brown AEX (2017) Predicting path from undulations for c. elegans
  using linear and nonlinear resistive force theory.
\newblock {\em Physical Biology} 14(2):025001.

\bibitem{Gray1955}
Gray J, Hancock GJ (1955) The {Propulsion} of {Sea}-{Urchin} {Spermatozoa}.
\newblock {\em Journal of Experimental Biology} 32(4):802--814.

\bibitem{Lauga2006}
Lauga E, DiLuzio WR, Whitesides GM, Stone HA (2006) Swimming in circles: Motion
  of bacteria near solid boundaries.
\newblock {\em Biophysical Journal} 90(2):400--412.

\bibitem{Liu2011}
Liu B, Powers TR, Breuer KS (2011) Force-free swimming of a model helical
  flagellum in viscoelastic fluids.
\newblock {\em Proceedings of the National Academy of Sciences}
  108(49):19516--19520.

\bibitem{Bayly2011}
Bayly P, et~al. (2011) Propulsive forces on the flagellum during locomotion of
  chlamydomonas reinhardtii.
\newblock {\em Biophysical Journal} 100(11):2716--2725.

\bibitem{Zhang2014}
Zhang T, Goldman DI (2014) The effectiveness of resistive force theory in
  granular locomotiona).
\newblock {\em Physics of Fluids} 26:101308.

\bibitem{Helms2019}
Helms SJ, et~al. (2019) {Modelling the ballistic-to-diffusive transition in
  nematode motility reveals variation in exploratory behaviour across species}.
\newblock {\em Journal of The Royal Society Interface} 16(157):20190174.

\bibitem{Stephens2010}
Stephens GJ, Johnson-Kerner B, Bialek W, Ryu WS (2010) {From modes to movement
  in the behavior of Caenorhabditis elegans.}
\newblock {\em PLoS One} 5(11):e13914.

\bibitem{Salvador2014}
Salvador LCM, Bartumeus F, Levin SA, Ryu WS (2014) {Mechanistic analysis of the
  search behaviour of \emph{Caenorhabditis elegans}}.
\newblock {\em Journal of The Royal Society Interface} 11(92):20131092.

\bibitem{Alves2017}
Alves LGA, et~al. (2017) Long-range correlations and fractal dynamics in c.
  elegans: Changes with aging and stress.
\newblock {\em Phys. Rev. E} 96(2):022417.

\bibitem{Artale1997}
Artale V, Boffetta G, Celani A, Cencini M, Vulpiani A (1997) {Dispersion of
  passive tracers in closed basins: Beyond the diffusion coefficient}.
\newblock {\em Physics of Fluids} 9(11):3162--3171.

\bibitem{Froyland2005}
Froyland G (2005) {Statistically optimal almost-invariant sets}.
\newblock {\em Physica D: Nonlinear Phenomena} 200(3-4):205--219.

\bibitem{VanKampen1992}
van Kampen N (1992) {\em {S}tochastic {P}rocesses in {P}hysics and
  {C}hemistry}.
\newblock (Elsevier Science Publishers, Amsterdam).

\bibitem{Ma2013}
Ma T, Bollt EM (2013) Relatively coherent sets as a hierarchical partition
  method.
\newblock {\em International Journal of Bifurcation and Chaos} 23(07):1330026.

\bibitem{Schutte2001}
Sch{\"u}tte C, Huisinga W, Deuflhard P (2001) Transfer operator approach to
  conformational dynamics in biomolecular systems in {\em Ergodic Theory,
  Analysis, and Efficient Simulation of Dynamical Systems}, ed.{} Fiedler B.
\newblock (Springer Berlin Heidelberg, Berlin, Heidelberg), pp. 191--223.

\bibitem{Bollt2013}
Bollt EM, Santitissadeekorn N (2013) {\em Applied and computational measurable
  dynamics}.
\newblock (Society for Industrial and Applied Mathematics, Philadelphia, United
  States).

\bibitem{Dellnitz1999}
Dellnitz M, Junge O (1999) {On the Approximation of Complicated Dynamical
  Behavior}.
\newblock {\em SIAM Journal on Numerical Analysis} 36(2):491--515.

\bibitem{Froyland2003}
Froyland G, Dellnitz M (2003) Detecting and locating near-optimal
  almost-invariant sets and cycles.
\newblock {\em SIAM Journal on Scientific Computing} 24(6):1839--1863.

\bibitem{Brown2013}
Brown AEX, Yemini EI, Grundy LJ, Jucikas T, Schafer WR (2013) {A dictionary of
  behavioral motifs reveals clusters of genes affecting Caenorhabditis elegans
  locomotion}.
\newblock {\em Proceedings of the National Academy of Sciences of the United
  States of America} 110(2):791--796.

\bibitem{Reddy2022}
Reddy G, et~al. (2022) A lexical approach for identifying behavioural action
  sequences.
\newblock {\em PLOS Computational Biology} 18(1):1--29.

\bibitem{KAPLAN2020}
Kaplan HS, {Salazar Thula} O, Khoss N, Zimmer M (2020) Nested neuronal dynamics
  orchestrate a behavioral hierarchy across timescales.
\newblock {\em Neuron} 105(3):562--576.e9.

\bibitem{Fujiwara2002}
Fujiwara M, Sengupta P, McIntire SL (2002) {Regulation of body size and
  behavioral state of C. elegans by sensory perception and the EGL-4
  cGMP-dependent protein kinase.}
\newblock {\em Neuron} 36(6):1091--102.

\bibitem{BenArous2009}
{Ben Arous} J, Laffont S, Chatenay D (2009) {Molecular and sensory basis of a
  food related two-state behavior in C. elegans}.
\newblock {\em PLoS ONE} 4(10):1--8.

\bibitem{Flavell2013}
Flavell SW, et~al. (2013) {Serotonin and the neuropeptide PDF initiate and
  extend opposing behavioral states in C. elegans}.
\newblock {\em Cell} 154(5):1023--1035.

\bibitem{Hills2004}
Hills T, Brockie PJ, Maricq AV (2004) {Dopamine and Glutamate Control
  Area-Restricted Search Behavior in Caenorhabditis elegans}.
\newblock {\em Journal of Neuroscience} 24(5):1217--1225.

\bibitem{Calhoun2014}
Calhoun AJ, Chalasani SH, Sharpee TO (2014) Maximally informative foraging by
  \textit{Caenorhabditis elegans}.
\newblock {\em eLife} 3:e04220.

\bibitem{Kareiva1987}
Kareiva P, Odell G (1987) Swarms of predators exhibit "preytaxis" if individual
  predators use area-restricted search.
\newblock {\em The American Naturalist} 130(2):233--270.

\bibitem{Bell1990}
Bell WJ (1990) Searching behavior patterns in insects.
\newblock {\em Annual Review of Entomology} 35(1):447--467.

\bibitem{Calhoun2015}
Calhoun AJ, et~al. (2015) {Neural mechanisms for evaluating environmental
  variability in caenorhabditis elegans}.
\newblock {\em Neuron} 86(2):428--441.

\bibitem{Hums2016}
Hums I, et~al. (2016) {Regulation of two motor patterns enables the gradual
  adjustment of locomotion strategy in caenorhabditis elegans}.
\newblock {\em eLife} 5(MAY2016):1--36.

\bibitem{Flavell2020}
Flavell SW, Raizen DM, You YJ (2020) {Behavioral States}.
\newblock {\em Genetics} 216(2):315--332.

\bibitem{Gray2004}
Gray JM, et~al. (2004) {Oxygen sensation and social feeding mediated by a C.
  elegans guanylate cyclase homologue}.
\newblock {\em Nature} 430(6997):317--322.

\bibitem{Neimark1959}
Neimark ED, Shuford EH (1959) Comparison of predictions and estimates in a
  probability learning situation.
\newblock {\em Journal of Experimental Psychology} 57(5):294--298.
\newblock Place: US Publisher: American Psychological Association.

\bibitem{Bullock1962}
Bullock DH, Bitterman ME (1962) Probability-matching in the pigeon.
\newblock {\em The American Journal of Psychology} 75(4):634--639.

\bibitem{Bitterman1958}
Bitterman ME, Wodinsky J, Candland DK (1958) Some comparative psychology.
\newblock {\em The American Journal of Psychology} 71(1):94--110.

\bibitem{Vulkan2000}
Vulkan N (2000) An economist’s perspective on probability matching.
\newblock {\em Journal of Economic Surveys} 14(1):101--118.

\bibitem{Newell2017}
Newell BR, Schulze C (2017) Probability matching. in {\em Cognitive illusions:
  {Intriguing} phenomena in thinking, judgment and memory, 2nd ed.}
\newblock (Routledge/Taylor \& Francis Group, New York, NY, US), pp. 62--78.

\bibitem{Saldana2022}
Saldana C, Claidière N, Fagot J, Smith K (2022) Probability matching is not
  the default decision making strategy in human and non-human primates.
\newblock {\em Scientific Reports} 12(1):13092.

\bibitem{Slivkins2019}
Slivkins A (2019) Introduction to multi-armed bandits.
\newblock {\em Found. Trends Mach. Learn.} 12(1–2):1–286.

\bibitem{Schwarz2015}
Schwarz RF, Branicky R, Grundy LJ, Schafer WR, Brown AEX (2015) Changes in
  postural syntax characterize sensory modulation and natural variation of c.
  elegans locomotion.
\newblock {\em PLOS Comput. Biol.} 11(8: e1004322):1--16.

\bibitem{Gomez-Marin2016}
Gomez-Marin A, Stephens GJ, Brown AEX (2016) {Hierarchical compression of
  \emph{Caenorhabditis elegans} locomotion reveals phenotypic differences in
  the organization of behaviour}.
\newblock {\em Journal of The Royal Society Interface} 13(121):20160466.

\bibitem{Berman2016}
Berman GJ, Bialek W, Shaevitz JW (2016) {Hierarchy and predictability in
  Drosophila behavior}.
\newblock {\em Proceedings of the National Academy of Sciences}
  104(51):20167--20172.

\bibitem{Gupta2019}
Gupta S, Gomez-Marin A (2019) {A context-free grammar for \emph{Caenorhabditis
  elegans} behavior, bioRxiv [Preprint] (2019).
  https://www.biorxiv.org/content/10.1101/708891v1 (accessed 19 October 2023)}.

\bibitem{BERG1972}
Berg HC, Brown DA (1972) {Chemotaxis in Escherichia coli analysed by
  Three-dimensional Tracking}.
\newblock {\em Nature} 239(5374):500--504.

\bibitem{Roberts2016}
Roberts WM, et~al. (2016) {A stochastic neuronal model predicts random search
  behaviors at multiple spatial scales in C. elegans}.
\newblock {\em Elife} 5:e12572.

\bibitem{Kato2015}
Kato S, et~al. (2015) {Global brain dynamics embed the motor command sequence
  of Caenorhabditis elegans}.
\newblock {\em Cell} 163:1--14.

\bibitem{Venkatachalam2016}
Venkatachalam V, et~al. (2016) Pan-neuronal imaging in roaming caenorhabditis
  elegans.
\newblock {\em Proceedings of the National Academy of Sciences of the United
  States of America} 113(8):1082--1088.

\bibitem{Hallinen2021}
Hallinen KM, et~al. (2021) Decoding locomotion from population neural activity
  in moving \textit{C. elegans}.
\newblock {\em eLife} 10:e66135.

\bibitem{Yemini2021}
Yemini E, et~al. (2021) {NeuroPAL}: {A} {Multicolor} {Atlas} for
  {Whole}-{Brain} {Neuronal} {Identification} in {C}. elegans.
\newblock {\em Cell} 184(1):272--288.e11.

\bibitem{Atanas2023}
Atanas AA, et~al. (2023) Brain-wide representations of behavior spanning
  multiple timescales and states in {C}.elegans.
\newblock {\em Cell} 186(19):4134--4151.e31.

\bibitem{Tinbergen1963}
Tinbergen N (1963) {On aims and methods of Ethology}.
\newblock {\em Zeitschrift f{\"{u}}r Tierpsychologie} 20:410--433.

\bibitem{Gallagher2013}
Gallagher T, Bjorness T, Greene R, You YJ, Avery L (2013) {The Geometry of
  Locomotive Behavioral States in C. elegans}.
\newblock {\em PLoS ONE} 8(3: e59865).

\bibitem{Jurgens2021}
Jurgens AM, Crutchfield JP (2021) Shannon {Entropy} {Rate} of {Hidden} {Markov}
  {Processes}.
\newblock {\em Journal of Statistical Physics} 183(2):32.

\bibitem{Shalizi2001}
Shalizi CR, Crutchfield JP (2001) Computational {Mechanics}: {Pattern} and
  {Prediction}, {Structure} and {Simplicity}.
\newblock {\em Journal of Statistical Physics} 104(3):817--879.

\bibitem{Crutchfield1989}
Crutchfield JP, Young K (1989) Inferring statistical complexity.
\newblock {\em Phys. Rev. Lett.} 63(2):105--108.

\bibitem{Lin2021}
Lin YT, Tian Y, Livescu D, Anghel M (2021) Data-driven learning for the
  mori--zwanzig formalism: A generalization of the koopman learning framework.
\newblock {\em SIAM Journal on Applied Dynamical Systems} 20(4):2558--2601.

\bibitem{Gilani2021}
Gilani F, Giannakis D, Harlim J (2021) Kernel-based prediction of non-markovian
  time series.
\newblock {\em Physica D: Nonlinear Phenomena} 418:132829.

\bibitem{costa2023fluctuating}
Costa AC, Vergassola M (2023) Fluctuating landscapes and heavy tails in animal
  behavior.

\bibitem{Froyland2010}
Froyland G, Lloyd S, Santitissadeekorn N (2010) {Coherent sets for
  nonautonomous dynamical systems}.
\newblock {\em Physica D: Nonlinear Phenomena} 239(16):1527--1541.

\bibitem{Froyland2010b}
Froyland G, Santitissadeekorn N, Monahan A (2010) Transport in time-dependent
  dynamical systems: Finite-time coherent sets.
\newblock {\em Chaos: An Interdisciplinary Journal of Nonlinear Science}
  20(4):043116.

\bibitem{Wang2015}
Wang H, Schütte C (2015) Building markov state models for periodically driven
  non-equilibrium systems.
\newblock {\em Journal of Chemical Theory and Computation} 11(4):1819--1831.
\newblock PMID: 26889513.

\bibitem{Koltai2016}
Koltai P, Ciccotti G, Schütte C (2016) On metastability and markov state
  models for non-stationary molecular dynamics.
\newblock {\em The Journal of Chemical Physics} 145(17):174103.

\bibitem{Koltai2018}
Koltai P, Wu H, No{\'{e}} F, Sch{\"{u}}tte C (2018) {Optimal data-driven
  estimation of generalized markov state models for non-equilibrium dynamics}.
\newblock {\em Computation} 6(1):1--23.

\bibitem{Fackeldey2019}
Fackeldey K, et~al. (2019) {From metastable to coherent sets -
  Time-discretization schemes}.
\newblock {\em Chaos} 29(1):012101.

\bibitem{Martineau2020}
Martineau CN, Brown AEX, Laurent P (2020) {Multidimensional phenotyping
  predicts lifespan and quantifies health in Caenorhabditis elegans}.
\newblock {\em PLOS Computational Biology} 16(7):1--14.

\bibitem{Ikeda2020}
Ikeda Y, et~al. (2020) C. elegans episodic swimming is driven by multifractal
  kinetics.
\newblock {\em Scientific Reports} 10(1):14775.

\bibitem{Le2020}
Le KN, et~al. (2020) An automated platform to monitor long-term behavior and
  healthspan in caenorhabditis elegans under precise environmental control.
\newblock {\em Communications Biology} 3(1):297.

\bibitem{Thompson1933}
Thompson WR (1933) On the likelihood that one unknown probability exceeds
  another in view of the evidence of two samples.
\newblock {\em Biometrika} 25(3/4):285--294.

\bibitem{MaBouDi2020}
MaBouDi H, Marshall JAR, Barron AB (2020) Honeybees solve a multi-comparison
  ranking task by probability matching.
\newblock {\em Proceedings of the Royal Society B: Biological Sciences}
  287(1934):20201525.

\bibitem{Beron2022}
Beron CC, Neufeld SQ, Linderman SW, Sabatini BL (2022) Mice exhibit stochastic
  and efficient action switching during probabilistic decision making.
\newblock {\em Proceedings of the National Academy of Sciences}
  119(15):e2113961119.

\bibitem{Schulz2020}
Schulz E, Franklin NT, Gershman SJ (2020) Finding structure in multi-armed
  bandits.
\newblock {\em Cognitive Psychology} 119:101261.

\bibitem{Niv2002}
Niv Y, Joel D, Meilijson I, Ruppin E (2002) Evolution of reinforcement learning
  in uncertain environments: A simple explanation for complex foraging
  behaviors.
\newblock {\em Adaptive Behavior} 10(1):5--24.

\bibitem{Green2010}
Green CS, Benson C, Kersten D, Schrater P (2010) Alterations in choice behavior
  by manipulations of world model.
\newblock {\em Proceedings of the National Academy of Sciences}
  107(37):16401--16406.

\bibitem{Braeckman2002}
Braeckman BP, Houthoofd K, Vanfleteren JR (2002) Assessing metabolic activity
  in aging caenorhabditis elegans: concepts and controversies.
\newblock {\em Aging Cell} 1(2):82--88.

\bibitem{Boulin2012}
Boulin T, Hobert O (2012) From genes to function: the c. elegans genetic
  toolbox.
\newblock {\em WIREs Developmental Biology} 1(1):114--137.

\bibitem{manuscript_data_doi}
Costa AC, Ahamed T, Jordan D, Stephens GJ (2023) Dataset
  (\url{https://doi.org/10.34740/KAGGLE/DS/3882219}).

\bibitem{Sulston1974}
Sulston JE, Brenner S (1974) {The DNA of Caenorhabditis elegans.}
\newblock {\em Genetics} 77(1):95--104.

\bibitem{scikit-learn}
Pedregosa F, et~al. (2011) Scikit-learn: Machine learning in {P}ython.
\newblock {\em Journal of Machine Learning Research} 12:2825--2830.

\bibitem{ARPACK}
Lehoucq RB, Sorensen DC, Yang C (1998) {\em ARPACK Users' Guide}.
\newblock (Society for Industrial and Applied Mathematics).

\bibitem{Scipy}
Jones E, Oliphant T, Peterson P, et~al. (2001--) {SciPy}: Open source
  scientific tools for {Python}.

\bibitem{Rabets2014}
Rabets Y, Backholm M, Dalnoki-Veress K, Ryu WS (2014) Direct measurements of
  drag forces in c. elegans crawling locomotion.
\newblock {\em Biophysical journal} 107 8:1980--1987.

\bibitem{Welch1967}
Welch P (1967) The use of fast fourier transform for the estimation of power
  spectra: A method based on time averaging over short, modified periodograms.
\newblock {\em IEEE Transactions on Audio and Electroacoustics} 15(2):70--73.

\end{thebibliography}

\clearpage

\onecolumngrid

\setcounter{figure}{0}
\setcounter{page}{1}

\makeatletter
\renewcommand{\theequation}{S\arabic{equation}}
\renewcommand{\thefigure}{S\arabic{figure}}

\section*{SI Appendix: A Markovian dynamics for \emph{C. elegans} behavior across scales}

\section*{Materials and Methods}

\noindent{\bf Software and data availability:} Code for reproducing our results is publicly available:
\url{https://github.com/AntonioCCosta/markov_worm/}. Data can be found in \cite{manuscript_data_doi}. 

\medskip

\noindent{\bf \emph{C. elegans} foraging dataset: } We used a previously-analyzed dataset \cite{Stephens2008}, in which young-adult N2-strain \textit{C.~elegans} were originally imaged at $f=32\,{\rm Hz}$ with a video tracking microscope on a food-free plate and then downsampled to $f=16\,{\rm Hz}$ to speed-up the process of resolving coiled postures \cite{Broekmans2016}. Worms were grown at $20{^\circ C}$ under standard conditions \cite{Sulston1974}. Before imaging, worms were removed from bacteria-strewn agar plates using a platinum worm pick, and rinsed from \textit{E.~coli} by letting them swim for $1\, {\rm min}$ in NGM (Nematode Growth Medium) buffer. They were then transferred to an assay plate ($9\, {\rm cm}$ Petri dish) that contained a copper ring ($5.1\, \rm{cm}$ inner diameter) pressed into the agar surface, preventing the worm from reaching the side of the plate. Recording started approximately $5\, \rm{min}$ after the transfer and lasted for $2100\, \rm{s}$, for a total of $T = 33600\,\text{frames}$. Each frame is converted into a 5-dimensional ``eigenworm'' representation $\vec{a}(t)$ by projecting the local tangent angles along the worm's centerline onto an ``eigenworm'' basis \cite{Stephens2008}, Fig.\,1(a).

\medskip

\begin{figure*}[ht!]
\begin{center}
\includegraphics[scale=1]{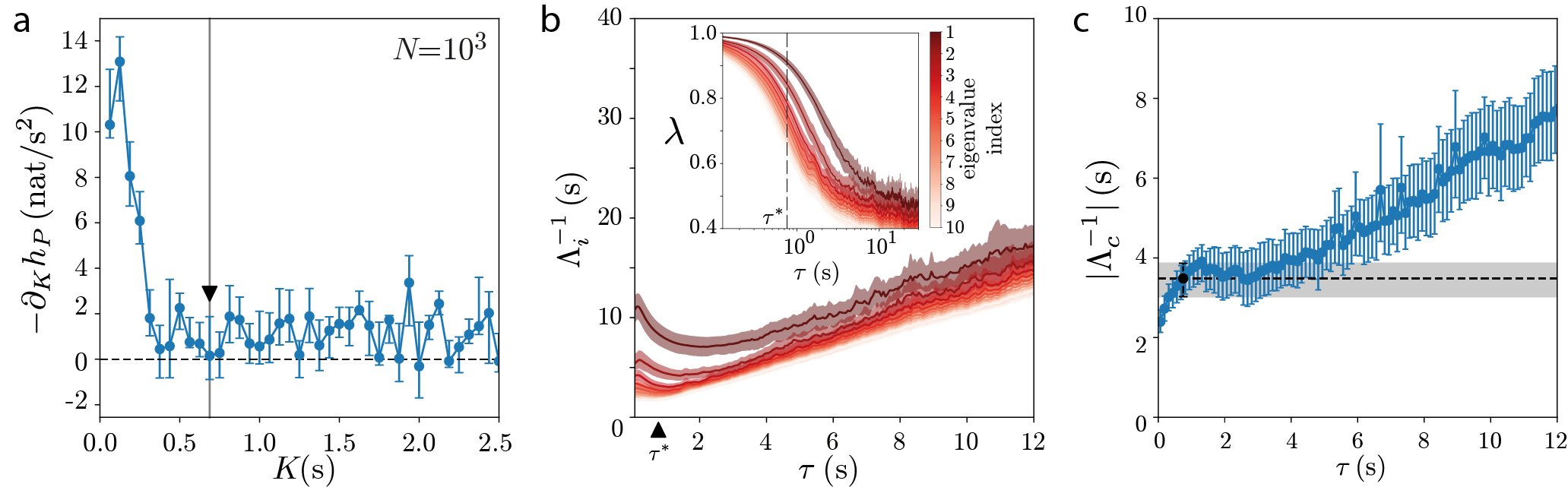}
\caption{{ \bf Details of the Markovian approximation of the \emph{C. elegans} posture dynamics through maximally predictive states.} (a) - Change in short-time entropy rate as a function of delays $K$ for $N=1000$ partitions. The entropy rates reaches a plateau after $K\gtrsim 0.5\,\text{s}$ and we choose $K^* = 11\,\text{frames} = 0.6875\,\text{s}$. Error bars represent 95\% confidence intervals bootstrapped across worms.
(b) - The ten largest relaxation timescales of the reversibilized transition matrix as a function of transition time $\tau$, and corresponding eigenvalues (inset). For $\tau\rightarrow \delta t$ the transition matrix is nearly the identity matrix (within $\tau$ most transitions occur within each partition), resulting in nearly degenerate eigenvalues close to 1 and an overestimation of the relaxation timescales of the reversibilized dynamics. On the other hand, when $\tau \gtrsim 5\,\text{s}$ the dynamics is mostly mixed, meaning that the transition matrix is composed of near copies of the steady-state distribution. In this regime, the eigenvalues of $P_r$, $\lambda_i$, become approximately constant, and therefore $\Lambda_i^{-1}(\tau) = -\tau/\log \lambda_i(\tau)$ grows linearly with $\tau$. Between these two regimes, the relaxation timescales are approximately constant, and this robustness to $\tau$ is indicative of Markovian dynamics. We choose $\tau^* = 0.75\,\text{s}$ as the shortest $\tau^*$ consistent with Markovian dynamics. Error bars are 95\% confidence intervals bootstrapped across worms.
(c) - The reversibilized transition matrix provides an optimal partition into almost invariant sets (see Section {\bf \uppercase{Coarse-graining behavior through ensemble dynamics}} for details), but the resulting kinetics does not necessarily capture the underlying dynamics. In fact, the obtained relaxation times are only an upper bound to the true relaxation timescales of the locally irreversible dynamics. To directly probe the Markovianity of the underlying slow dynamics, we estimate the relaxation times for the non-reversibilized coarse-grained transition matrix, which should not change with $\tau$ when the dynamics is Markovian. We approximate the slow relaxation dynamics by using the metastable states to build a two-state, coarse-grained Markov chain $P_c$, which necessarily has only real eigenvalues $\lambda_c\in\mathbb{R}$. The corresponding relaxation time is then obtained through $|\Lambda_c^{-1}| = -\tau/\log\lambda_c(\tau)$. 
In general, we find that the regime in which $|\hat{\Lambda}_2^{-1}|$ from $P_r$ is constant (b) overlaps with regime in which $|\Lambda_c^{-1}|$ is also constant. In addition, while $|\Lambda_2^{-1}|(\tau^*)$ from $P_r$ overestimates the expected $|\Lambda^{-1}|$ from ``run'' and ``pirouette' transition rates, Fig.\,4(b), the timescales obtained from $P_c$, $|\Lambda_c^{-1}|(\tau^*)= 3.48(3.03, 3.86)\,\text{s}$ are comparable to the ones estimated from the entire Markov chain in Fig.\,2(a) and accurately predict the hopping dynamics. Error bars are $95\%$ confidence intervals bootstrapped across worms.
}
\label{fig_S1}
\end{center}
\end{figure*}

\noindent{\bf Maximally predictive states:} Given the measurement time series, $\vec{a}(t)$, with $t\in\{\delta t,\ldots,T \delta t\}$ and $\vec{a}\in \mathbb{R}^5$, we build a trajectory matrix by stacking $K$ time-shifted copies of $\vec{a}$, yielding a $(T-K)\times Kd$ matrix $X_K$. For each $K$, we partition the candidate state space and estimate the entropy rate of the associated Markov chain (see below). We choose $K^*$ such that $\partial_K h(K^*) \sim 0$, which defined $X_K^*$ as the maximally predictive states \cite{Costa2023}, Fig.\,\ref{fig_S1}(a).

\medskip

\noindent{\bf State space partitioning:} We partition the state space constructed from the ensemble of worms into $N$ Voronoi cells, $s_i, i\in\{1,\ldots,N\}$, through k-means clustering with a k-means++ initialization using scikit-learn \cite{scikit-learn}.

\medskip

\noindent{\bf Transition matrix estimation:} We build a finite dimensional approximation of the Perron-Frobenius operator using an Ulam-Galerkin discretization \cite{Bollt2013}. In practice, given T observations, a set of $N$ partitions, and a transition time $\tau$, we compute

\begin{equation}
    C_{ij}(\tau) = \sum_{t=0}^{T-\tau}\zeta_i(X_{K^*}(t))\zeta_j(X_{K^*}(t+\tau)), \nonumber
\end{equation}

\noindent where $\zeta_i(x)$ are the Ulam basis functions, which are characteristic functions
\begin{equation*}
    \zeta_i(x) = \begin{cases}
    1, & \text{for $x \in s_i$}\\ 
    0, & \text{otherwise}
    \end{cases}
\end{equation*}
set by the k-means clustering. The maximum likelihood estimator of the transition matrix is obtained by simply row normalizing the count matrix,

\begin{equation}\label{eq:Pij}
    P_{ij}(\tau) = \frac{C_{ij}(\tau)}{\sum_j C_{ij}(\tau)}, \nonumber
\end{equation}

\noindent which yields an approximation of the Perron-Frobenius operator. 
\medskip

\noindent{\bf Invariant density estimation:} Given a transition matrix $P$, the invariant density is obtained through the left eigenvector of the non-degenerate eigenvalue 1 of $P$, $\pi P = \pi$: $\pi_i$ is the probability of finding the system in a partition $s_i$.

\medskip

\noindent{\bf Short-time entropy rate estimation:} Given a number of partitions $N$ and a sampling time scale $\tau=\delta t$, we estimate the Markov transition matrix $P$ and the corresponding invariant density $\pi$ as detailed above and compute the short-time entropy rate as,

\begin{align}\label{eq:h_Markov}
    h = -\frac{1}{\delta t}\sum_{ij}\pi_i P_{ij}\log P_{ij}.
\end{align}
To obtain error bars in Fig.\,1, S1(a), we estimate a transition matrix for each worm using its symbolic sequence, and then estimate its corresponding entropy rate.

\medskip

\noindent{\bf Two-dimensional UMAP embedding:} We use the UMAP embedding \cite{McInnes2018} as a tool to visualize the maximally predictive states of \emph{C. elegans} posture dynamics. In a nutshell, the UMAP algorithm searches for a low dimensional representation of the data that preserves its topological structure. We use a publicly available implementation of the algorithm found in \href{https://github.com/lmcinnes/umap}{\tt https://github.com/lmcinnes/umap}, within which we chose the Chebyshev distance metric to compute distances in the high-dimensional space, {\tt n\_neighbors}=50 nearest neighbors and {\tt min\_dist}=0.05 as the minimum distance.

\medskip

\noindent {\bf Matrix diagonalization:} The high dimensionality and the sparsity of the  transition matrices for large $N$ results in numerical errors when using a naive estimator for the full spectrum of eigenvalues. In addition, since we are interested in the longest lived dynamics, we focus on finding only the $n_\text{modes}$ largest magnitude real eigenvalues using the ARPACK \cite{ARPACK} algorithm.

\medskip

\noindent {\bf Choice of transition time $\tau^*$:} We choose $\tau^*$ such that the resulting Markovian dynamics approximate the long-term behavior of the system accurately, as in \cite{Costa2023}. In practice, we find the shortest transition time scale after which the inferred implied relaxation times reach a plateau, Fig.\,\ref{fig_S1}(b,c). For $\tau$ too short, the approximation of the operator yields a transition matrix that is nearly identity (due to the finite size of the partitions and too short transition time), which results in degenerate eigenvalues close to $\lambda\sim 1$: an artifact of the discretization and not reflective of the underlying dynamics. For $\tau$ too large, the transition probabilities become indistinguishable from noisy estimates of invariant density, which results in a single surviving eigenvalue $\lambda_1=1$ while the remaining eigenvalues converge to a noise floor resulting from a finite sampling of the invariant density. Between such regimes, we find a region with the largest time scale separation which also corresponds to the regime for which the longest relaxation times, Eq.\,(2), are robust to the choice of $\tau$, Fig.\ref{fig_S1}(b,c). To obtain a noise floor (horizontal line in Fig.\,4(a-right)), we shuffle the symbolic sequence, reestimate the transition matrix, and compute its first nontrivial eigenvalue. In the limit of infinite data, this shuffle contains only one surviving nonzero eigenvalue corresponding to the steady-state distribution (infinite relaxation time). The observation that the second largest eigenvalue is nonzero even in the shuffle is due to finite-size effects that result in small deviations from the invariant density. For further discussion see \cite{Costa2023}. To estimate error bars in Figs.\,4,S1, we estimate the eigenvalues of transition matrices obtained for each worm.

\medskip

\noindent {\bf Cross-validation experiments:} We chose the number of partitions $N^*$ so as to capture as much finescale detail of the dynamics without inducing finite-size effects on the estimates of the entropy rate, Fig.\,1. To further attest that this choice results in generalizable models, we additionally performed cross-validation experiments as follows. We split each worm dataset into 10 segments of 3.5 minutes each, and randomly select 3 segments as a test set and the remaining 7 segments as a training set. We repeat this process over 50 random shuffles, estimating a transition matrix from the training set and making predictions over the unseen test data. In Fig.\,S4(a-c), we simulate the test data with the model parameters estimated in the training data, and compare such simulations against the data bootstrapping over 50 random reshuffles of train-test sets.

\medskip
\noindent{\bf \emph{C. elegans} posture simulations:} At each iteration, we sample from the conditional distribution given by the Markov chain inferred for each worm $P^w(s_j(t+\tau^*)|s_i(t))$ to generate a symbolic sequence sampled on a timescale $\tau^*$. We then randomly sample a state space point $X_{K^*}$ within the partition $s_i$, and unfold it to obtain a sequence of postures $\vec{a}_{t:t+K^*}$ at each $\tau^*$. We can thus generate artificial posture time series with the same duration as the experimental time series (35 minutes), but with a missing frame every $\tau^*$ frames (the gap between $K^*$ and $\tau^*$), which we interpolate across using a cubic spline with scipy's {\tt interpolate} package  \cite{Scipy}, and smooth with a cubic polynomial and a window size of 11 frames using the {\tt signal.savgol\_filter} package from Scipy \cite{Scipy}. We then take the simulated $\vec{a}(t)$ time series and transform it back to the tangent angles at each body segment $\theta_i(t)$ using the ``eigenworms'' \cite{Stephens2008}.

\medskip

\noindent{\bf Estimating the rate of reversals, dorsal and ventral turn events:} Reversal events where identified as segments in which the absolute value of the worms' overall curvature $\gamma (t) = \sum_i \theta_i(t)$ was $|\gamma|<3\times 10^{-4}\,\text{rad}$ and the body wave phase velocity $\omega(t) = -\frac{1}{2\pi}\frac{d}{dt}\left[\text{tan}^{-1}(a_2(t)/a_1(t))\right]$ \cite{Stephens2008} was $\omega<-0.2\,\text{cycles}\,\text{s}^{-1}$ for at least $0.5\,\text{s}$. Ventral and dorsal turns were identified as segments where the overall body curvature was either $\gamma<-3.5\times 10^{-4}\,\text{rad}$ or $\gamma>3.5\times 10^{-4}\,\text{rad}$, respectively, for at least $0.5\,\text{s}$. 

\medskip

\noindent{\bf Resistive force theory simulations:} We recover the rigid body motion from the tangent angle time series using linear resistive force theory, as in \cite{Keaveny2017}. We approximate the forces acting independently on each body segment as 
\begin{align*}
    \tilde{\bf F}_i(t) = \alpha_t \tilde{v}_i^t \hat{t}+\alpha_n \tilde{v}_i^n \hat{n}
\end{align*}
where $\tilde{v}_i^{t,n}$ are the tangent and normal components of the velocity at each segment $i$, which can be written in terms of the velocity and displacements measured after subtracting the overall rigid body motion,

\begin{align*}
    \tilde{\bf v}_i(t) = {\bf v}_i(t) + \tilde{\bf V}(t) + \tilde{\bf \Omega}(t)\times \Delta{\bf x}_i(t).
\end{align*}
Then, by imposing a zero net-force and net-torque condition at each frame,
\begin{align*}
    \sum_i \tilde{\bf F}_i &= 0 \\
    \sum_i \tilde{\bf F}_i \times \Delta{\bf x}_i &= 0,
\end{align*}
we obtain a system of linear equations that for a given $\alpha = \alpha_n/\alpha_t$ can be solved for the components of the worm's velocity $\tilde{\bf V}(t)$ and angular velocity $\tilde{\bf \Omega}(t)$ \cite{Keaveny2017}. From these we can integrate the path taken by the worm's body to obtain a reconstructed $\tilde{\bf x}_\text{CM}(t)$.

\begin{figure*}
\begin{center}
\includegraphics[scale=1]{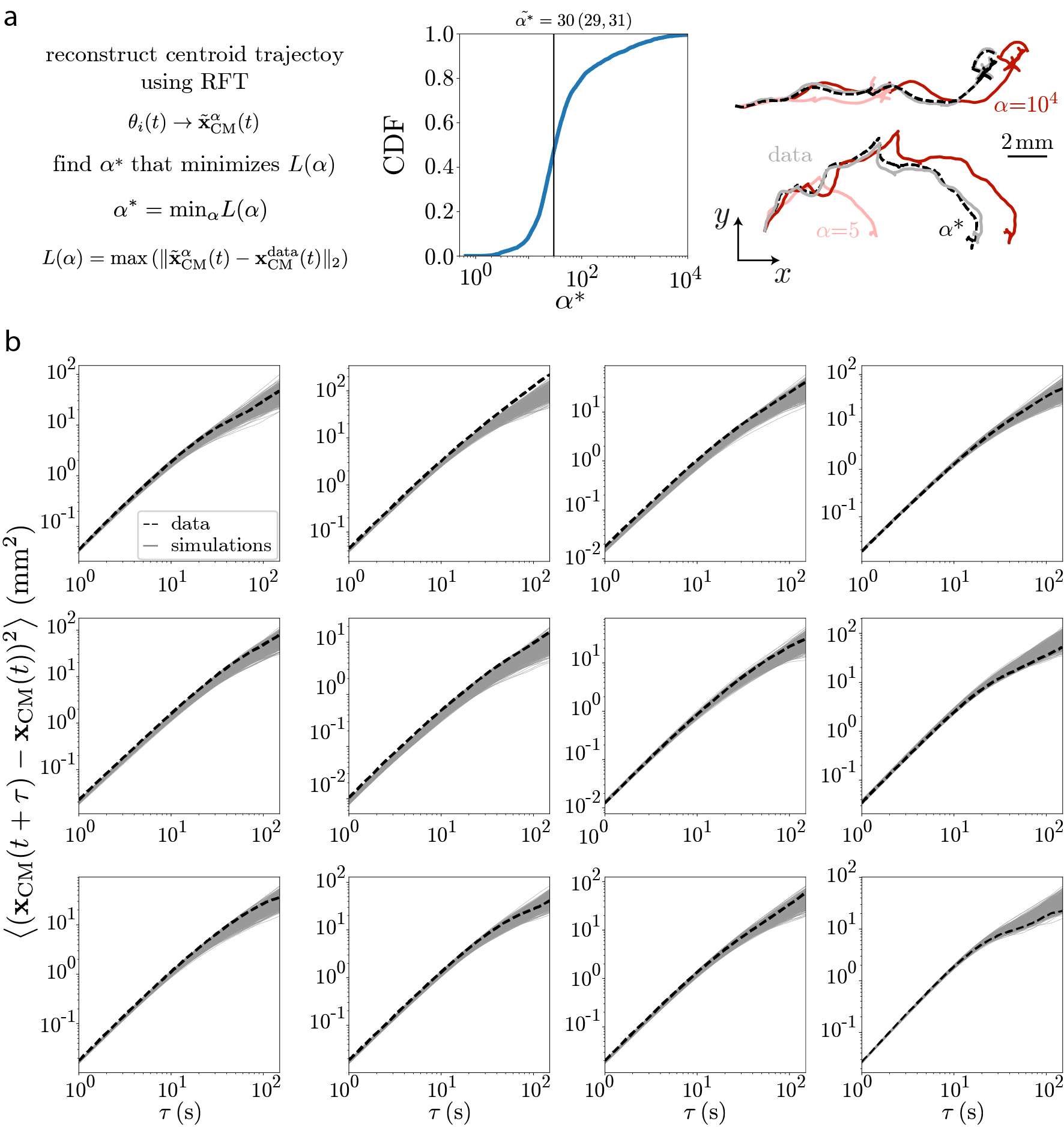}
\caption{{ \bf Details of the posture-to-path simulations.}
(a) Optimizing the free parameter $\alpha = \alpha_n/\alpha_t$ in RFT (see Methods for details). We define a loss function $L(\alpha)$ as the maximum distance between the real worm trajectory and a reconstructed one, and sample randomly from $100\,\text{s}$ segments. We the find the optimal $\alpha^*$ through the Nelder-Mead algorithm (see Methods). The resulting distribution of $\alpha^*$ is broad, as shown through the cumulative distribution function (CDF), with a median of $\tilde{\alpha^*} = 30\,(29,31)\,\text{s}^{-1}$. On the right, we display two example trajectory reconstructions for different values of $\alpha$. For $\alpha=5$ (comparable to experimental measures of \cite{Rabets2014}), RFT typically results in a substantial undershoot of the observed trajectories (as observed in \cite{Keaveny2017}). For the no-slip condition, $\alpha \gg 1$, the resulting trajectories overshoot the real worm trajectories, indicating that some degree of slip is needed to accurately predict worm trajectories. With values of $\alpha^*\approx 30$ we get an accurate reconstruction of the worm trajectories.
(b) Mean square displacements for the data (dashed line) of each worm, as well as for 1000 centroid trajectory simulations generated from symbolic sequences simulated with the Markov model (gray). By fitting a linear function in the interval $\tau\in[60,100]\,\text{s}$ we obtained the effective diffusivity estimate of Fig.\,4(b-right).
}
\label{fig_S_RFT}
\end{center}
\end{figure*}

We optimize the single free parameter $\alpha$ by comparing the reconstructed trajectories with the real worm trajectories ${\bf x}^\text{data}_\text{CM}$, Fig.\,\ref{fig_S_RFT}(a). In particular, we minimize the maximum distance between $100\,\text{s}$ trajectories randomly sampled from the dataset $L(\alpha) = \text{max}(\lVert \tilde{\bf x}^\alpha_\text{CM}(t) - {\bf x}^\text{data}_\text{CM}(t)\rVert_2),\, t\in [t_0,t_0+100\,\text{s}]$. To minimize $L(\alpha)$ we use the Nelder-Mead algorithm through the {\tt scipy.optimize} library of Scipy \cite{Scipy}. The software to translate posture into path can be found in \url{https://github.com/AntonioCCosta/markov_worm}, and follows closely the implementation of \cite{Keaveny2017}.

\medskip

\begin{figure*}[ht!]
\begin{center}
\includegraphics[scale=1]{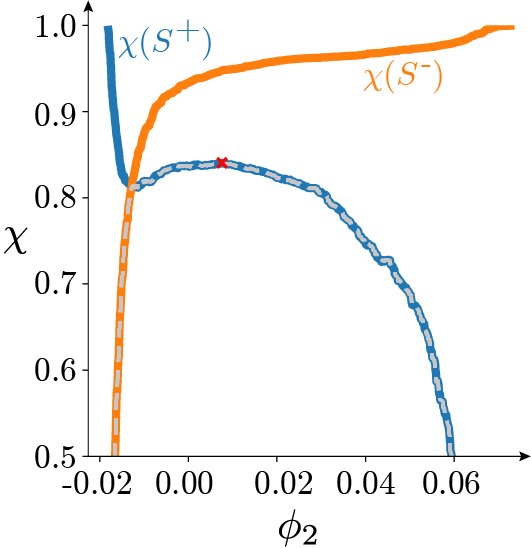}
\caption{{ \bf Coherence measure used to define the metastable states.} We define metastable states by maximizing the overall coherence, Eq.\,(\ref{Eq:coherence_min}), of the two macroscopic states obtained by partitioning the state-space along $\phi_2$ (see Methods for details). We here plot the coherence of each set (orange and blue), as well as the overall minima across sets $\chi$, Eq.\,(\ref{Eq:coherence_min}) (gray dashed line). The maximum of $\chi$ is highlighted with a red cross and indicates the value of $\phi_2$ that defines the metastable states, $\phi_2 = \phi_2^c$.
}
\label{fig_S_coherence}
\end{center}
\end{figure*}

\noindent{\bf Metastable states:} Metastable states correspond to collections of short-time movements that typically follow each other in time to give rise to stereotyped sequences. Leveraging our previous work \cite{Costa2023}, we search for metastable states along the slowest mode of the reversibilized dynamics \cite{Dellnitz1999}. As shown in \cite{Froyland2005}, the second eigenvector $\phi_2$ of a time-reversibilized transition matrix $P_r$ provides an \emph{optimal} subdivision of the state space into almost invariant sets. In practice, we use the ensemble of worms to estimate $P_r$ as

\begin{equation}
    P_r(\tau) = \frac{P(\tau)+P(-\tau)}{2},
\end{equation}\label{Eq:rev_P}
\noindent where, 
\begin{equation}
    P_{ij}(-\tau) = \frac{\pi_j P_{ji}(\tau)}{\pi_i} \nonumber
\end{equation}

\noindent is the stochastic matrix governing the time-reversal of the Markov chain. The first non-trivial ($\lambda<1$) right eigenvector of $P_r$, $\phi_2$, allows us to define macrostates as collections of microstates $s_i$,

\begin{equation*}
    S^+ (\phi_2^c) \coloneqq \bigcup_{i:\phi_2\geq \phi_2^c} s_i\,\\,\,S^- (\phi_2^c)\coloneqq \bigcup_{i:\phi_2<\phi_2^c} s_i,
\end{equation*}

\noindent where $\phi_2^c$ is a threshold that is chosen to maximize the metastability of a set. We measure the metastability of each set $S$ by estimating how much of the probability density remains in $S$ after a time scale $\tau$, 

\begin{equation*}\label{Eq:coherence}
    \chi_{\pi,\tau}(S) = \frac{\sum_{i,j\in S}\pi_i P_{ij}(\tau)}{\sum_{i \in S}  \pi_i}.
\end{equation*}
To estimate the overall measure of metastability across both sets $S^+$ and $S^-$, we define

\begin{equation}\label{Eq:coherence_min}
    \chi(\phi_2^c) = \min\left\{\chi_{\pi,\tau^*}(S^+),\chi_{\pi,\tau^*}(S^-)\right\}.
\end{equation}
which we maximize with respect to $\phi_2^c$. Metastable states are then defined with respect to the sign of $\phi_2 - \phi_2^c$. See \cite{Costa2023} for further details and applications to known dynamical systems. In Fig.\,\ref{fig_S_coherence} we show the overall coherence measure as a function of $\phi_2$ for the worm data.

\medskip

\noindent {\bf Operator-based state space subdivision:} We leverage the notion of relatively coherent sets \cite{Ma2013} to subdivide the state space. However, instead of subdividing both metastable state at each iteration $k$, we identify the state with the most measure $S^*_k$ and build a new transition matrix only with partitions belonging to that state,

\begin{equation*}
    P_{S^*_k}(\tau)= p(s_j(t+\tau)|s_i(t)),\, i,j\in S^*_k.
\end{equation*}

\noindent From $P_{S^*_k}$ we proceed as before: we compute the stationary distribution of $S^*_k$ through the first left eigenvector of $P_{S^*_k}$, $\pi^*_i$, build the corresponding reversibilized transition matrix $P_{r,S^*_k}$ and identify relatively metastable states through its first non-trivial eigenvector by maximizing Eq.\,(\ref{Eq:coherence_min}) where $\pi_i$ and $P_{ij}(\tau)$ are replaced by their relative counterparts $\pi^*_i$ and $P_{S^*_k}$.

\medskip 

\noindent{\bf Simulating posture-to-path within mesoscopic behavioral states:} To generate a centroid trajectory within a given state, we construct a transition matrix among the partitions corresponding to each of the mesoscopic states identified in Fig.\,2(c) using data from all worms. We then proceed as in Figs.\,2,3 to generate both posture time series and centroid trajectories. We first generate a symbolic sequence by sampling states according to the corresponding transition probability matrix $\hat{s}_j(t+\tau)\sim P_S(s_j|\hat{s}_i(t))$, $i,j \in S$. From the symbolic sequence, we then sample a time series segment $\vec{a}_{t:t+K^*}$ within each sampled partition, and use resistive force theory to translate the resulting $\theta(t)$ time series into locomotion. In this way, we can simulate posture and centroid trajectories for \emph{in silico} worms that are forced to remain within a particular mesoscopic behavioral state for an arbitrary amount of time.

\medskip

\noindent{\bf Probability of finding food as a function of distance and behavioral state:} We estimate the likelihood of finding food in a given radius $r$ by estimating the fraction of the area within a disc of radius $r$ covered by the worm's body during $100\,\text{s}$ trajectories, taking the worm's width to be $5\%$ of its length. We then normalize these area fractions by the total across states, obtaining the $p(\text{food}|r,\text{state})$ showed in Fig.\,5(b).

\clearpage

\begin{figure*}[ht!]
\begin{center}
\includegraphics[scale=1]{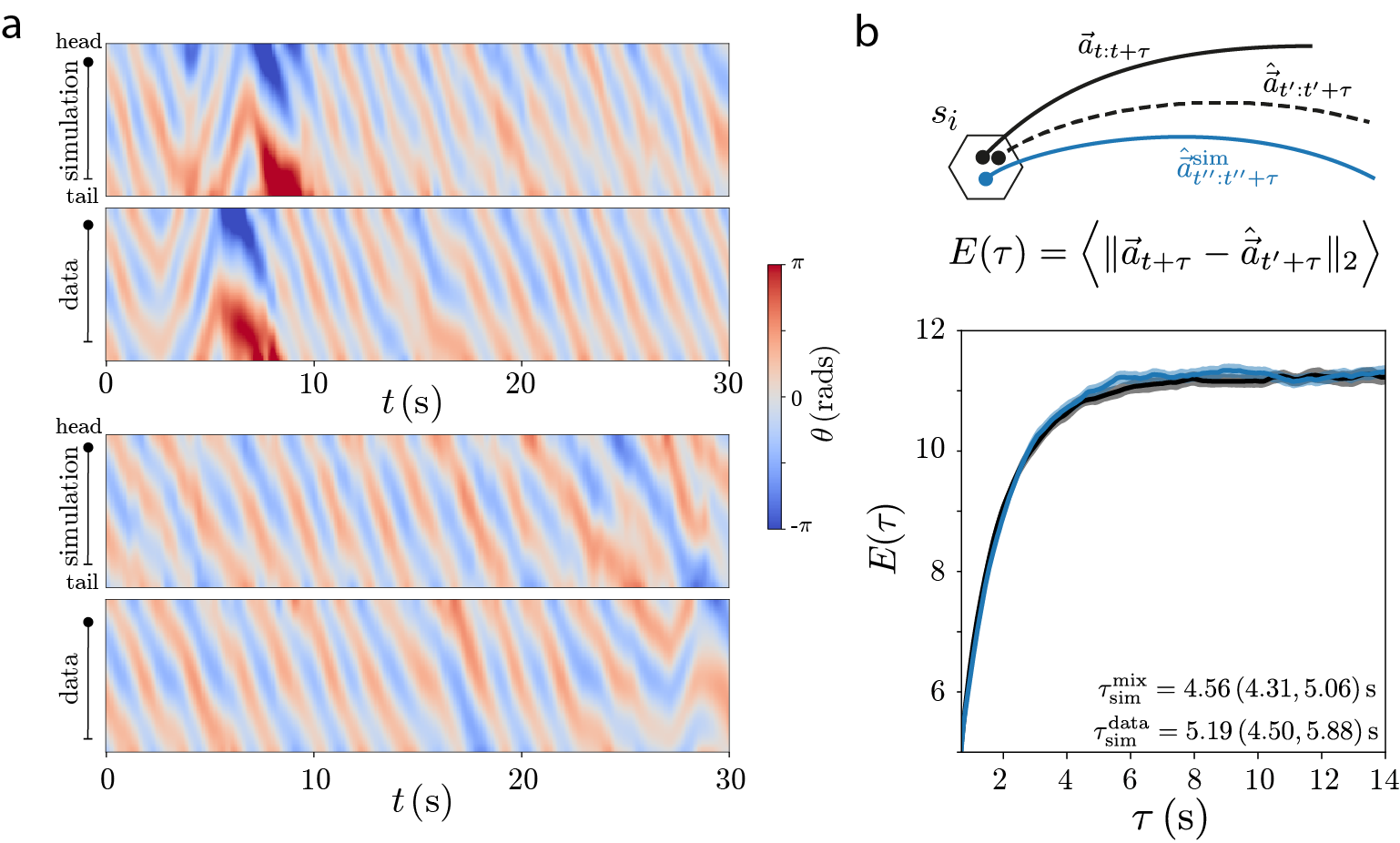}
\caption{{ \bf Details of the Markov chain simulations.} (a) - Two illustrative curvature vs. time plots comparing simulations with data. As expected due to the unpredictable nature of the dynamics \cite{Ahamed2021}, the quality of the predictions worsens as time progresses. Nonetheless, the structure of the dynamics is well preserved, making it hard to know \emph{a priori} which of the two time traces is the data and which one is a simulation. (b) - We assess the predictive power of the Markov model by estimating how prediction errors grow over time. We define $E(\tau)$ by estimating the average distance between two trajectories starting within the same partition: $t$ is chosen at random and $t'\neq t$ chosen such that $\hat{\vec{a}}_{t'}$ belongs to the same partition as $\hat{\vec{a}}_{t}$; the expectation value is then taken over multiple samples of $t$. We compare $E(\tau)$ estimated from simulations (blue) against sampling a trajectory from the data starting from the same partition (black). As summary statistics, we compute the time it takes before predictions completely mix, obtaining $\tau^\text{mix}_\text{data} = 5.19\,(4.50,5.88)\,\text{s}$ for the data, just slightly higher than $\tau^\text{mix}_\text{sim} = 4.56\,(4.31,5.06)\,\text{s}$ from simulations. In practice, we estimate the average distance between two randomly sampled points $e_\infty = \left\langle \lVert \vec{a}_i(t) - \vec{a}_i(t') \rVert_2 \right\rangle_{t,t'}$, and find the time it takes for $E(\tau)\leq 0.95 e_\infty$.
}
\label{fig_S_posture_predictions}
\end{center}
\end{figure*}

\begin{figure*}
\begin{center}
\includegraphics[scale=1]{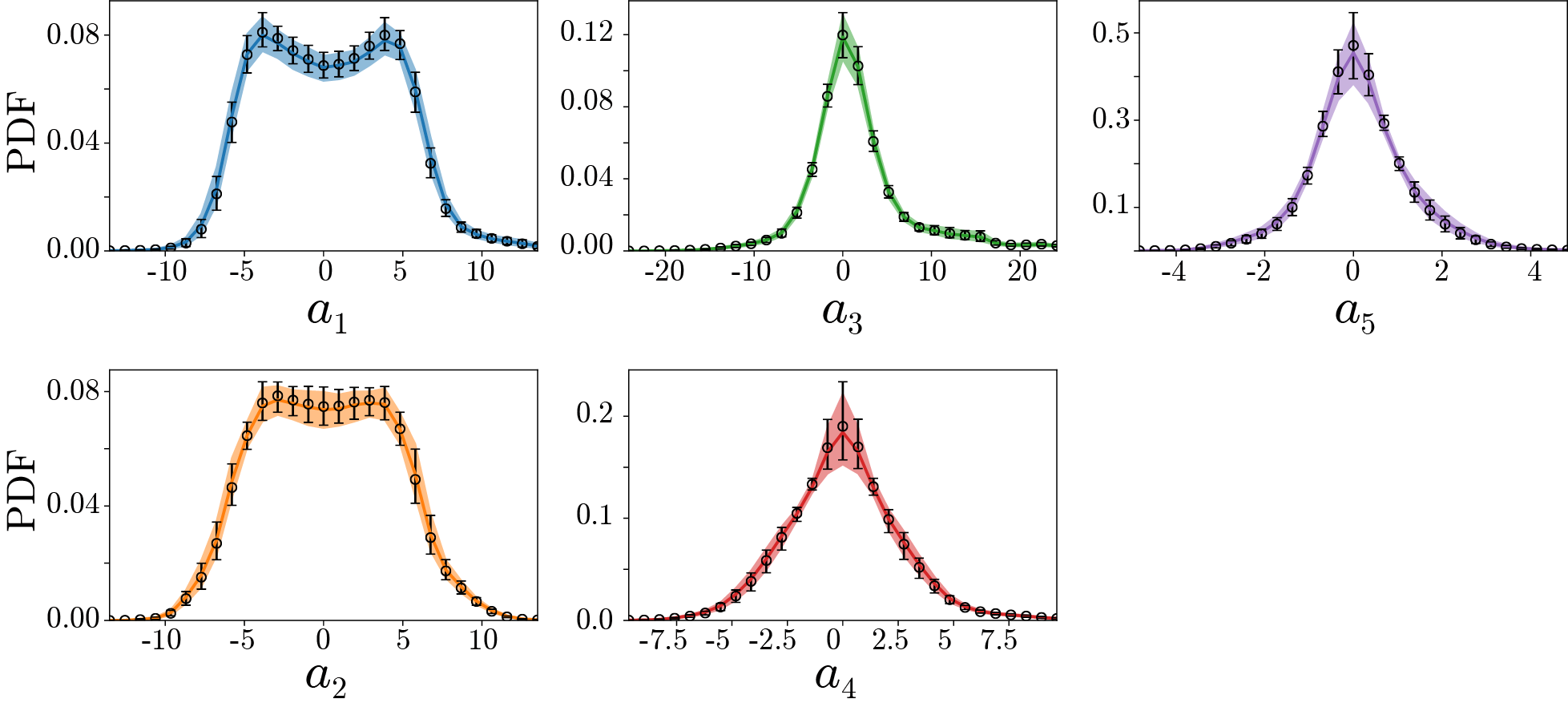}
\caption{
Probability Density Function (PDF) of the ``eigenworm'' coefficients show a tight agreement between the data (colors) and simulations (black error bars), indicating that the inferred dynamics capture the steady-state distribution. Error bars and shaded areas correspond to 95\% confidence intervals bootstrapped across worms.
}
\label{fig_S_posture_steady_state}
\end{center}
\end{figure*}

\begin{figure*}
\begin{center}
\includegraphics[scale=1]{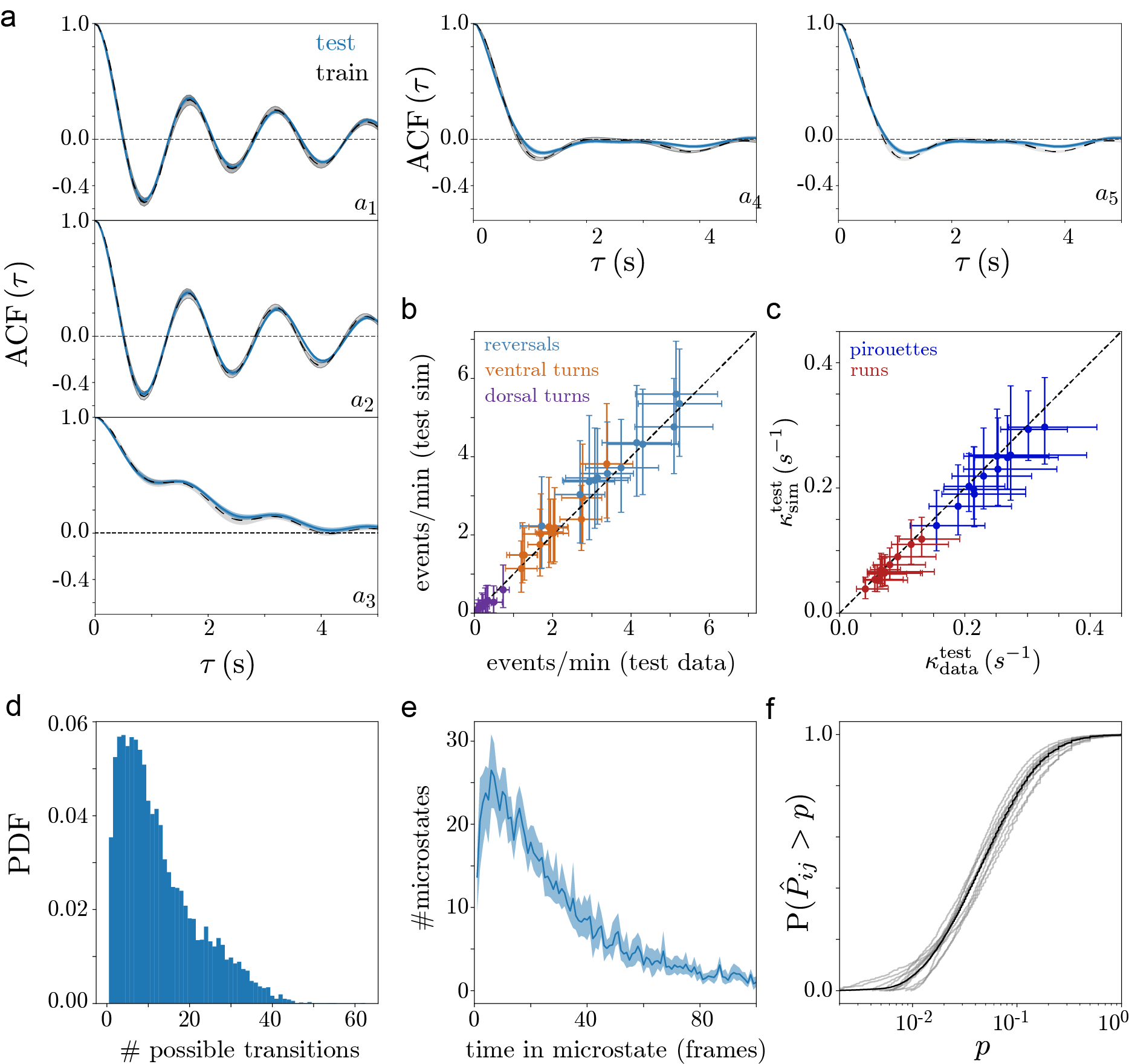}
\caption{{\bf Cross-validation of the Markov model and details of the transition matrix estimation.}
In Fig.\,1 we build a Markov model with $N^*=1000$ microstates to yield maximal expressivity before finite-size effects (which result in an underestimation of the entropy rate at large $N$) become evident. Here, to showcase how this choice of $N^*$ results in a robust and generalizable model for the posture dynamics of each worm, we perform a cross-validation experiment by splitting the data of each worm into randomly sampled 70-30 training-test sets, learn the $P_{ij}$ matrix using only 70\% of the data and predict on a 30\% unseen test set. We split the data into 10 segments (of $3.5\,\text{minutes}$ each), randomly choose 3 segments for a test set and the remaining 7 segments for a training set, and repeat this process over 50 random seeds (see Methods for details). (a) - Autocorrelation function of the ``eigenworm'' coefficient time series obtained from the test data (blue) and simulations of the test data obtained from a model inferred in a separate training dataset (black) for an example worm. Shaded areas represent 95\% confidence intervals obtained by bootstrapping across 50 random resampling of test and training datasets.
(b,c) - Cross-validation of the rate of different behavioral events (b) and the transition rates between run and pirouette states (c) for all worms. We plot them as in Figs.\,2(c,d), except that now we compare test set data with simulations obtained with a model built from a separate training set. Error bars correspond to 95\% confidence intervals bootstrapped across the sampled behavioral events in 50 random reshuffles of train-test sets.
(d) - To further showcase the estimation procedure, we assess how well transitions can be sampled. Despite the large number of microstates ($N^*=1000$), the number of possible transitions from each state is constrained by the fact that $\tau^*$ is much smaller than the mixing time, rendering the transitions local. Therefore, the transition matrix is extremely sparse, with most state visiting only $\approx 10$ other states resulting in a total of $\approx 10,000$ entries in the transition matrix. 
(e) - Histogram of the time spent in a microstate. While some microstates are visited rarely, the vast majority is visited at least 30 times. Error bars are bootstrapped across individual worms.
(f) - Cumulative distribution function of the non-zero entries to the transition matrix. Most of the measured transition probabilities are between $0.01 \lesssim P_{ij} \lesssim 0.5$. The black line represents the $P_{ij}$ estimated from the ensemble of worms, while each gray line represents the $P_{ij}$ for each individual worm.
}
\label{fig_S_crossval}
\end{center}
\end{figure*}

\begin{figure*}[ht!]
\begin{center}
\includegraphics[scale=1]{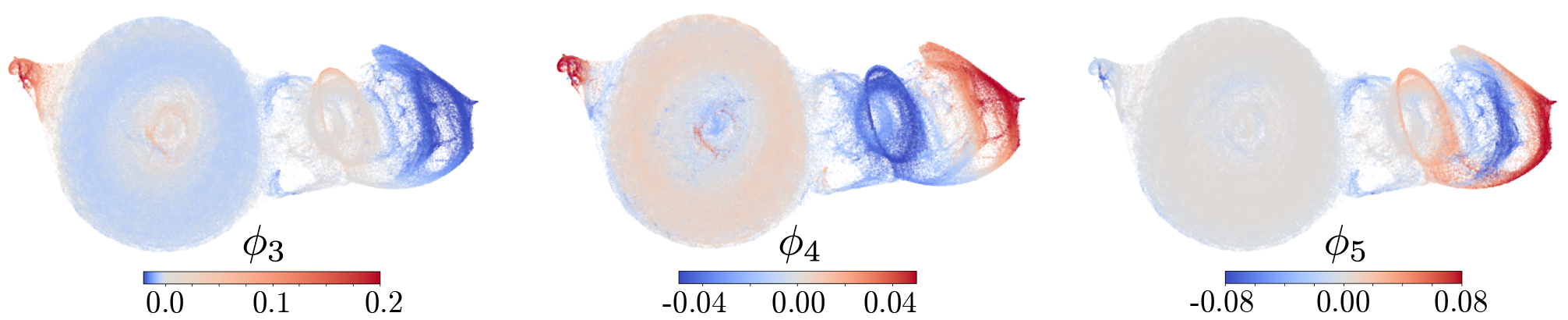}
\caption{{ \bf Beyond ``run and pirouette'', additional slow modes of \emph{C. elegans} behavior.}
While we have focused on $\phi_2$, there is also important information in the remaining long-lived eigenfunctions. We color code the worm's maximally predictive state space by the projection along the 3 following eigenvectors, $\{\phi_3,\phi_4,\phi_5\}$, which are organized according to their relaxation times $\Lambda_3^{-1}>\Lambda_4^{-1}>\Lambda_5^{-1}$. We see that $\phi_3$ differentiates dorsal and ventral turns, $\phi_4$ differentiates turning and reversals, and $\phi_5$ differentiates the compound motifs of shallow turns following a pause, from reversals that are followed by deep $\delta$-turns. Together with $\phi_2$ these modes provide a principled encoding of {\em C. elegans} off-food behavior across scales.
}
\label{fig_S_eigfun}
\end{center}
\end{figure*}

\begin{figure*}[ht!]
\begin{center}
\includegraphics[scale=1]{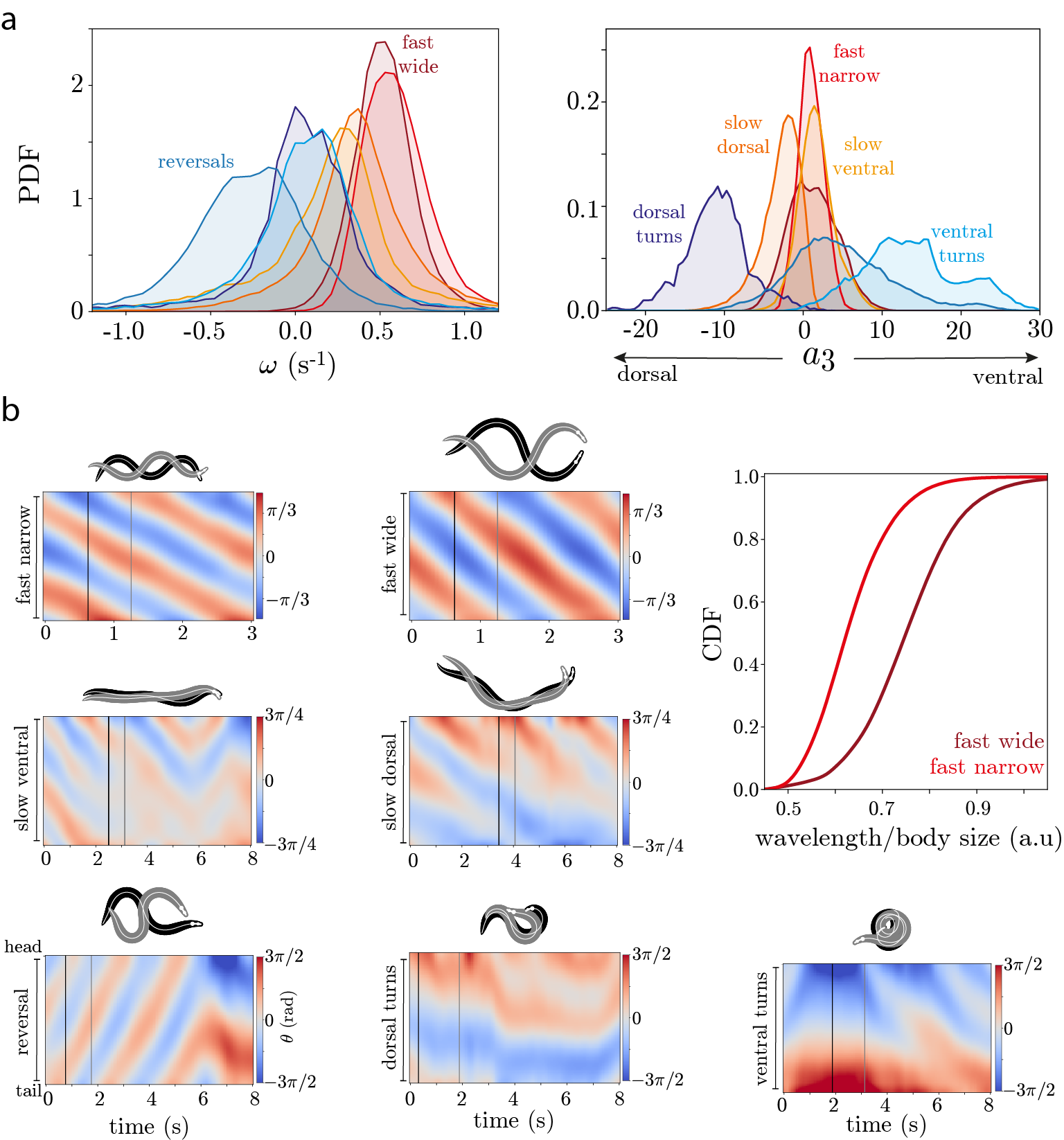}
\caption{{ \bf Characterization of the 7 mesoscopic states obtained through a top-down subdivision of the \emph{C. elegans} posture state space.} (a) - Probability Distribution Function (PDF) of the body wave phase velocity $\omega = -\frac{1}{2\pi}\frac{d}{dt}\left(\tan^{-1}(a_2/a_1)\right)$ (left) and the dorso-ventral turning amplitude (right), as measured through the third ``eigenworm'' coefficient $a_3$ \cite{Stephens2008,Broekmans2016}. The state labels were given based on these probability distributions. (b) Example local tangent angles $\theta_i$ as a function of time in each of the states, as well as illustrative postures sampled at different time points (black and gray vertical lines). The two fast states (top) correspond to distinct gaits with different wavelengths (right), while the slow states (middle) lack a coherent body wave traveling from head to tail. Instead, the dorsally-biased slow states exhibits short timescale head-casting behavior \cite{KAPLAN2020}, while the ventrally-biased state exhibits incoherent body motion with partial reversal and forward waves akin to a dwelling state \cite{Fujiwara2002}.
}
\label{fig_S_substates}
\end{center}
\end{figure*}

\begin{figure*}[ht!]
\begin{center}
\includegraphics[scale=1]{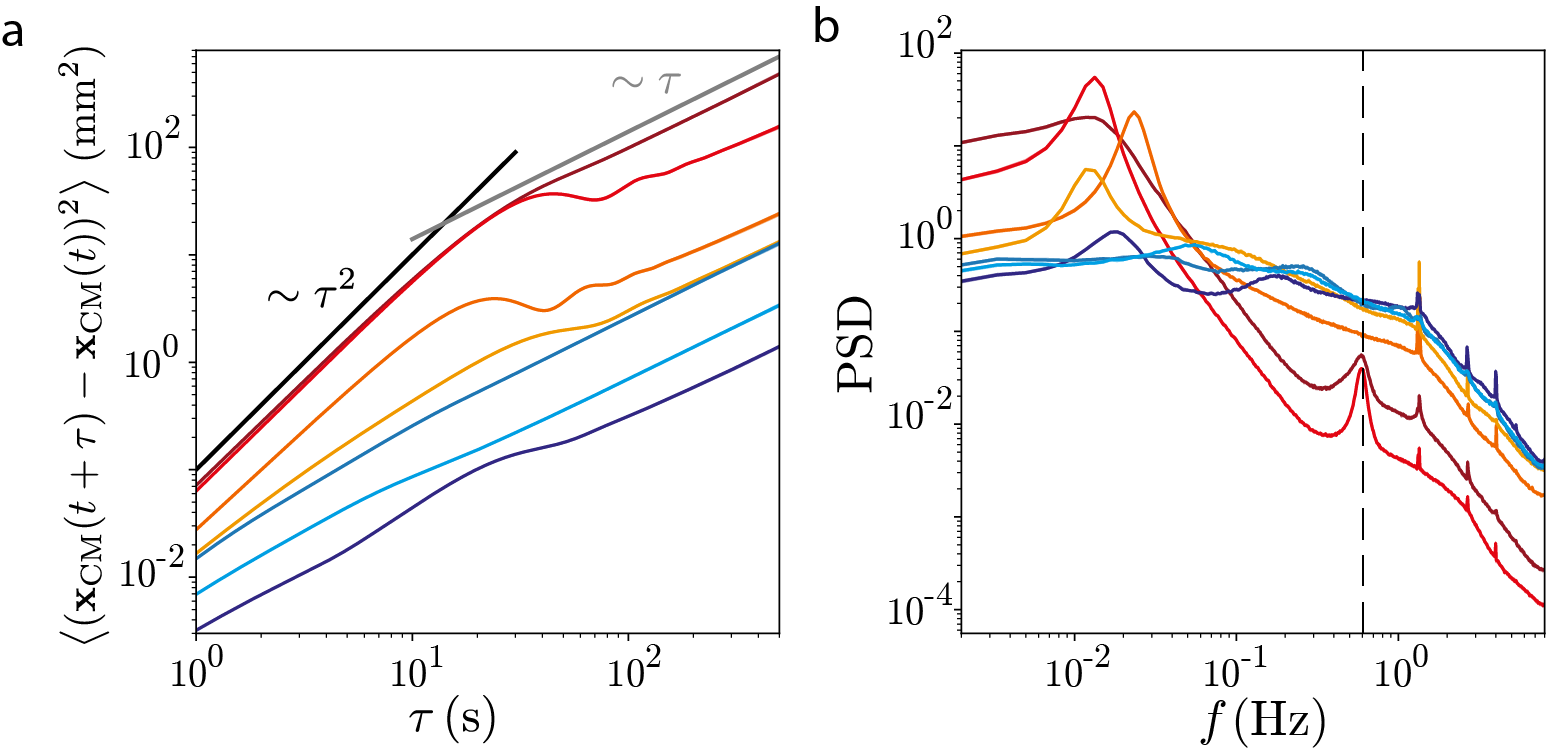}
\caption{{ \bf Statistical properties of the trajectories simulated for each of the 7 mesoscopic states obtained through a top-down subdivision of the \emph{C. elegans} posture state space.} (a) - Mean square displacement (MSD) of the center-of-mass trajectories for each of the 7 states revealed in Fig.\,4(c). Trajectories in the different states exhibit clearly distinct statistical properties: while ``run'' states generally exhibit a transition between super-diffusive behavior ($\text{MSD}\sim \tau^\beta,\, \beta>1$) at short times and diffusive (or sub-diffusive) behavior ($\text{MSD}\sim \tau^\beta,\, \beta\leq 1$) at large times, the ``pirouette''' states are mostly diffusive even at short times. In addition, some of the states exhibit non-trivial fluctuations in the MSD that result from the quasi-periodic loops observed in the trajectories shown in Fig.\,5(a). (b) - Power spectral density of the velocity bearing angle $\eta$, $\vec{v}_\text{CM} = v\cos(\eta) \,\hat{e}_x + v\sin(\eta) \,\hat{e}_y$, where $v = ||\vec{v}_\text{CM}||$. Besides the fast oscillations due to the body wave dynamics (the vertical dashed line represents the body wave phase velocity in the fast wide run state), some states also exhibit low frequency peaks due to the loopy nature of the trajectories, which recur on a time scale orders of magnitude longer than the body wave period. The power spectral density was estimated using Welch's method \cite{Welch1967} implemented through the {\tt signal.welch} package from Scipy \cite{Scipy} with a Hann window and $10\,\text{min}$ long trajectory segments. The error bars corresponding to 95\% confidence bootstrapped across 1000 simulations for each state are too small to show.
}
\label{fig_S_msd_psd_states}
\end{center}
\end{figure*}

\begin{figure*}[ht!]
\begin{center}
\includegraphics[scale=1]{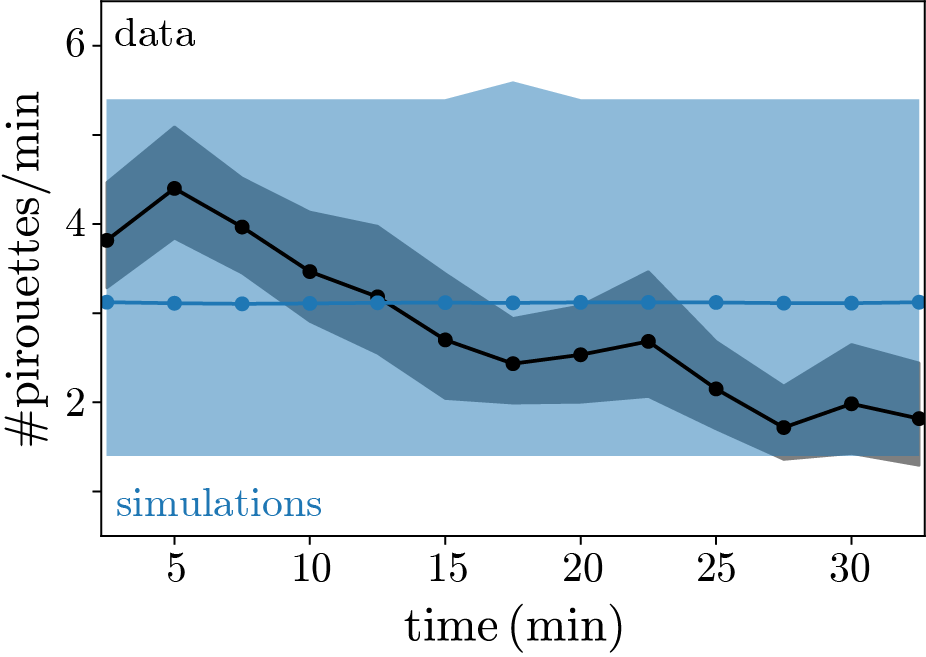}
\caption{{ \bf The rate of pirouettes changes with time on the food-free plate, a non-stationary dynamics not captured in our Markov model.} We estimate the rate of pirouettes as a function of time from the data (black), and find that it slowly decreases, reflecting a change in search strategy possibly as a result of updates to the animal's prior over the food distribution. Our simulations (blue) do not capture this slow change in the pirouette rate, as that would require an explicit time dependence in the transition probability matrix. Error bars correspond to 95\% confidence bootstrapped across the 12 worms in our dataset. To estimate the rate of pirouettes, we coarse-grain the dynamics into ``runs'' and ``pirouettes'' as in Fig.\,4, and find the number of pirouette events per minute in sliding $5\,\text{min}$ windows with $2.5\,\text{min}$ overlap. Since the sampling time of the Markov dynamics is $\tau^*$, we discard pirouette events with a duration shorter than $\tau^*$ from this analysis.
}
\label{fig_S_pir_per_min}
\end{center}
\end{figure*}

\clearpage

\noindent {\bf Movie S1:  \href{https://antonioccosta.github.io/download/postures_sim_vs_data.mp4}{(Download here)}} The simulated posture dynamics is virtually indistinguishable from the real worm data. We simulate a worm using the Markov model procedure of Fig.\,2, and compare its posture dynamics with data starting from the same microstate. We show the curvature over time (left) for real (bottom) and simulated (top) worms, which are \emph{a priori} indistinguishable from each other. We also show the corresponding RFT reconstructed skeletons (right), after subtracting the centroid position.  

\medskip

\noindent {\bf Movie S2: \href{https://antonioccosta.github.io/download/combined_traj.mp4}{(Download here)}} Illustration of the posture to path simulations. We simulate a worm using the Markov model procedure of Fig.\,2, and translate its posture dynamics into movement using the resistive force theory approach of Fig.\,3.

\end{document}


\title{Appendix}

\maketitle

\section*{Predictive information for state space reconstruction}

In real-world complex systems, individual measurements, even if high-dimensional, rarely capture the full set of variables that constitute the state space, which is Markovian and maximally predictive of the future.  Analyses based solely on such measurements might induce apparent complexity, simply due to the fact that important fine-scale predictive information is missing.  For example, measuring only the instantaneous position $\vec{x}(t)$ of a pendulum provides incomplete predictive information, as from a picture of a pendulum one cannot say whether the pendulum is moving in an upswing or a downswing at that specific point in time. Newtonian dynamics dictates that the position evolves according to a second-order ordinary differential equation (ODE), and thus we need to know the instantaneous velocity of the pendulum $\vec{v}(t)$ in order to be able to predict the future position $\vec{x}(t+\delta t)$. Therefore, measuring only the position $X = {\vec{x}(t)}$ yields an incomplete non-Markovian description of the system, as the instantaneous position is insufficient to predict the next time step. In contrast, defining a position-momentum ``phase space'' yields a Markovian state-space in the expanded $X = (\vec{x},\vec{v})$ space, for which the second-order ODE becomes a set of first-order ODEs. An alternative way to capture such additional predictive information, is to include temporal information into the definition of state. For example, defining $X = (\vec{x}(t),\vec{x}(t+\delta t))$, $0< \delta t\ll 1$, also yields a maximally predictive representation, as knowing the immediate past position also allow us to determine the future: this is analogous to having access to a short video of the pendulum, instead of just a snapshot. This intuition is captured mathematically through delay embedding theorems \cite{Takens1981,Stark1999,Stark2003}, which show that it is possible to reconstruct a system's state by expanding the measurement in time through time-delays. Typically, the goal of state-space reconstruction is to obtain a topologically equivalent representation from which to infer geometrically-invariant ergodic properties of the dynamics, such as dimensions or Lyapunov exponents. Interestingly, such quantities can guide the search of a model class, since they can reveal symmetries, the degree in which the dynamics is dissipative, its inherent dimension, etc. Embedding theorems are general, in the sense that they cover a wide-range of dynamical systems that are commonly used to model natural phenomena (including deterministic and stochastic dynamics), and therefore their applicability is in principle general. In practice however, it is often challenging to find the number of time delays and sampling timescales that accurately reconstruct the state-space.

Consider a set of incomplete measurements of an unknown dynamical system, $\vec{y}(t) = M(\vec{x}(t))$, where $M$ is a measurement function mapping the underlying state space $\vec{x}(t)\in\mathbb{R}^D$ into our measurements $\vec{y}(t)\in\mathbb{R}^d$, for which typically $d<D$, Fig.~1(b-left). We expand the putative state by adding $K-1$ time delays to the measurement time series, sampled on a time scale $\delta t$, yielding a candidate state space $X_K(t) \in \mathbb{R}^{d\times K}$. Much like the movement of the pendulum, in order to find a maximally-predictive state-space reconstruction we will leverage past information to narrow down our prediction of the immediate future. This intuition is naturally captured by the concept of entropy rate, which essentially measures how spread out is the distribution of future states given the current state. In practice, we bin the continuous state-space $X = \{X_K(0),X_K(\tau),\ldots,X_K(T)$ into a discrete partitioned space with $N$ voronoi cells, $s_X =\{s_0,s_{\tau},\ldots,s_T\}$ where $s_i \in \mathbb{I}^N$, and estimate the entropy rate as,

\begin{equation}
h_K(N,\tau) = -\frac{1}{\tau}\sum_{ij} \pi_i p\left(s_j(t+\tau)|s_i(t)\right)\log p\left(s_j(t+\tau)|s_i(t)\right),
\end{equation}
where $p\left(s_j(t+\tau)|s_i(t)\right)$ is the conditional probability of the future state $s_j(t+\tau)$ given the current states $s_i(t)$, $\tau$ is the transition time scale and $\pi$ is the invariant measure of the dynamics. Notably, $h_K(N,\tau)$ is a non-decreasing function of the number of partitions $\partial_N h_K(N,\tau)\geq 0$. The behavior of $h_K(N,\tau)$ with $N$, or equivalently, with a typical length scale $\epsilon \propto 1/N$, is indicative of different classes of dynamics \cite{Gaspard1993}. For example, stochastic dynamics possesses information on all length scales, and so $h_K(\infty,\tau)=\infty$, whereas the fractal nature of deterministic chaotic systems yields a typical length scale below which the entropy stops changing: $h_K(N^* = 1/\epsilon^*,\tau)=h_K(\infty,\tau) $. In general however, finite-size effects yield an underestimation of the entropy. With the aim of preserving as much information as possible in our state-space discretization, we set the number of partitions as the largest $N$ after which the entropy rate stops increasing, Fig.~1(b-middle). Therefore, we maximize the entropy with respect to the number of partitions, obtaining $h_K(\tau) = \max_N h_K(N,\tau)$.

The conditional probabilities define a row-stochastic matrix $P_{ij}(\tau)$, which evolves the state space densities by a time $\tau$,
\begin{equation}
p(s_j(t+\tau))=P_{ij}(\tau) p(s_i(t)),
\label{eq:Pshort}
\end{equation}
\noindent where we sum over repeated indices (Einstein's summation convention). The transition matrix $P_{ij}(\tau)$ built from a discretization of the state-space is an example of an approximated {\em transfer operator} $\mathcal{P_\tau}$, Eq.\,(\ref{eq:PF_operator}) \cite{Bollt2013}, Fig.~1(b-right), which the central object we use to coarse-grain the dynamics and identify metastable states. Thus, leveraging the entropy rate we can probe whether our transfer operator representation of the reconstructed dynamics is maximally predictive, unifying the transfer operator approximation and state-space reconstruction under a single quantity.

Our goal is to find the number of time delays $K^*$ that minimizes the entropy rate, maximizing predictive information  \cite{Bialek2001} $I_\text{pred} = \left\langle\log\frac{p(x_\text{future}|x_\text{past})}{p(x_\text{future})}\right\rangle \sim H_K(N,\tau)-h_K(N,\tau)$.  In general, we observe that as we include past information into the definition o

The transfer operator dynamics solely depends on the topology of the state space trajectories, which are guaranteed to be preserved by a state space embedding \cite{Takens1981,Stark2003}. In that sense, $\mathcal{P}_\tau$ is in principle exactly preserved, and maintains all the properties of the underlying dynamics. Therefore, given an appropriate number of time delays $K^*$, the local entropy rate of the reconstructed dynamics will be equivalent that of the underlying phase space $h_K(N,\tau)\sim\hat{h}_K(N,\tau)$, where $\hat{h}_K(N,\tau)$ is the local entropy rate computed from the underlying state-space dynamics. In contrast, the set of non-linear equations of motion driving the dynamics of the reconstructed state $\dot{X}=F(X)$, which are for instance required to obtain estimates of the local Jacobian, are more sensitive to the detailed geometric properties of the reconstructed space (such as dimensions, metric, etc.), making it non-trivial to accurately approximate the underlying dynamics $\dot{x} = f(x)$, Fig.~1(a,b-right). The transfer operator formalism is therefore complementary to trajectory based approaches, providing a means to study large scale properties of the dynamics while being robust to the precise geometric properties of the reconstructed state.

When $K$ is too short, $\delta h_{P_N}(K)$ will be large, meaning that a large amount of information is required in order to make an accurate prediction. In systems with finite range correlations, there is a $K^*$ for which $\delta h_{P,N}(K^*) = 0$, in which $K^*$ corresponds to the amount of memory sitting in the measurement time series. Our state space reconstruction seeks $K^*$ such that $h_{P_N}(K)$ is minimized, i.e.,
\begin{equation}
    \partial_K h_{P_N}(K^*) = 0,
\end{equation}
\noindent which corresponds to maximizing the predictive information.

\subsection*{Limitations of the state-space reconstruction framework}

Consider a set of incomplete measurements of an unknown dynamical system, $\vec{y} = M(\vec{x})$, where $M$ is a measurement function mapping the underlying state space $\vec{x}\in\mathbb{R}^D$ into our measurements $\vec{y}\in\mathbb{R}^d$, for which typically $d<D$, Fig.~1(b-left). We expand the putative state by adding $K-1$ time delays to the measurement time series, sampled on a time scale $\delta t$, yielding a candidate state space $X_K \in \mathbb{R}^{d\times K}$. We quantitatively characterize the unpredictability of the candidate state space through the entropy rate of the symbolic dynamics resulting from partitioning each $K\times d$-dimensional putative state space into $N$ Voronoi cells through clustering (Methods).  With a partitioning,  the reconstructed dynamics are encoded as a row-stochastic transition probability matrix $P$ which evolves a state-space density $p$ by a time $\delta t$, 
\begin{equation}
p_i(t+\delta t)=P_{ji}p_j(t),
\label{eq:Pshort}
\end{equation}
\noindent where we sum over repeated indices (Einstein's summation convention). The entropy rate of the source is approached by estimating the entropy rate of the associated Markov chain for increasing values of $K$,

\begin{equation}\label{eq:h_p}
    h_{P_N}(K) =  -\sum_{ij} \pi_{i} P_{ij} \text{log}P_{ij}
\end{equation}

\noindent where $\pi$ is the estimated stationary distribution of the Markov chain $P$\footnote{For simplicity, we use a slight abuse of notation: unlike the previous $\pi$ from $\mathcal{P}_\tau$, here $\pi\in\mathbb{R}^N$ is the finite approximation of the invariant density.}. The Markov approximation of the entropies provides an estimate of the conditional entropies between discrete states $s$, $\left\langle - \text{log}\left[p(s_j|s_i)\right] \right\rangle$, where $i,j \in \{1,\ldots, N\}$. Each discrete state contains a population of delay vectors $\vec{y}^K = \{\vec{y}_i,\ldots,\vec{y}_{i+K-1}\}$, and therefore the entropy of the Markov chain provides an estimate of the sequence entropy of the time series, 

\begin{align}
    h_{P_N}(K) &  \approx \left\langle -\text{log}\left[ p_N(\vec{y}_{i+K}|\vec{y}_i\ldots \vec{y}_{i+K-1} \right] \right\rangle  \nonumber \\ 
    & = H_{K}(N)-H_{K-1}(N) \nonumber\\
    & = h_K(N) \nonumber
\end{align}
\noindent where $H_{K}(N)$ is the entropy of the $K$-gram symbolic sequence built by discretizing the $\vec{y}$ space into $N$ partitions. Note that the entropy rate is a non-decreasing function on the number of partitions, N, which is equivalent to a typical state-space scale $\epsilon \sim 1/N$. Importantly, we seek to preserve as much information as possible in the discretized state space, setting the number of partitions as the largest $N$ after which the entropy rate stops increasing due to finite-size effects Fig.~1(b-middle). This yields the maximum entropy rate  with respect to the number of partitions and thus approximates the Kolmogorov-Sinai (KS) limit of the entropy rate \cite{Cohen1985,Gaspard1993}. Besides the KS entropy, other quantities such as information or correlation dimensions can in principle be obtained by studying the scaling of the measure with the partition size  \cite{Farmer1983,Grassberger1983,Cohen1985}.

With increasing K and an appropriate partition, we approach the entropy rate of the source from above so that the difference,
 \begin{equation}
    \delta h_{P,N}(K) = h_{P_N}(K-1)- h_{P_N}(K)\geq 0, \nonumber
\end{equation}
\noindent is a non-increasing function of $K$. $\delta h_{P,N}(K)$ has been used to define measures of forecasting complexity in dynamical systems \cite{Grassberger1986} and is the amount of information that has to be kept in the $K-1$ time delays for an accurate forecast of the next time step. When $K$ is too short, $\delta h_{P_N}(K)$ will be large, meaning that a large amount of information is required in order to make an accurate prediction. In systems with finite range correlations, there is a $K^*$ for which $\delta h_{P,N}(K^*) = 0$, in which $K^*$ corresponds to the amount of memory sitting in the measurement time series. Our state space reconstruction seeks $K^*$ such that $h_{P_N}(K)$ is minimized, i.e.,
\begin{equation}
    \partial_K h_{P_N}(K^*) = 0,
\end{equation}
\noindent which corresponds to maximizing the predictive information. 
We recall the general definition of predictive information \cite{Bialek2001}, $I_\text{pred} = \left\langle\log\frac{p(x_\text{future}|x_\text{past})}{p(x_\text{future})}\right\rangle$. \noindent Defining $x_\text{past}$ as the first $K-1$ time steps in the time series and  $x_\text{future}$ as the $K$th time step, 
\begin{align}
    I_\text{pred}& = \langle -\log p(\vec{y}_{i+K}) \rangle - \langle -\log p(\vec{y}_{i+K}|\vec{y}_i\ldots \vec{y}_{i+K-1}) \rangle. \nonumber \\ 
    & = H(\vec{y}) - h_K, \nonumber
\end{align}
and thus with respect to a partition into $N$ states, we have
\begin{equation}
    I_\text{pred}(N) = H_1(N) - h_{P_N}(K).
\label{Eq:Ipred}    
\end{equation}

\noindent $I_\text{pred}(N)$ thus quantifies how the uncertainty of our predictions reduces when we also know the transition probabilities and not just the steady-state distribution. The predictive information is maximized when $h_{P_N}(K)$ is minimized, which for a system with finite correlations is attained when $\partial_K h_{P_N}(K^*) = 0$.

Given a reconstructed state $X_{K^*}$, the transition matrix $P_{ij}$ (defined with timestep $\tau=\delta t$), is an example of an approximated {\em transfer operator} $\mathcal{P_\tau}$, Eq.\,(\ref{eq:PF_operator}) \cite{Bollt2013}, Fig.~1(b-right). The transfer operator dynamics solely depends on the topology of the state space trajectories, which are guaranteed to be preserved by a state space embedding \cite{Takens1981,Stark2003}. In that sense, $\mathcal{P}_\tau$ is in principle exactly preserved, and maintains all the properties of the underlying dynamics. In contrast, the set of non-linear equations of motion driving the dynamics of the reconstructed state $\dot{X}=F(X)$, which are for instance required to obtain estimates of the local Jacobian, strongly depend on the geometric properties of the space (such as dimensions, metric, etc.), making it non-trivial to accurately approximate the underlying dynamics $\dot{x} = f(x)$, Fig.~1(a,b-right). The transfer operator formalism is therefore complementary to trajectory based approaches, providing a means to study large scale properties of the dynamics while being robust to the precise geometric properties of the reconstructed state.

\clearpage

\bibliography{Bibliography}